\documentclass[usegraphicx,usenatbib]{mn2e}

\usepackage[total={17.8cm,24.0cm},centering]{geometry}
\usepackage{times,deluxetable}

\newcommand{\apj}{ApJ}           
\newcommand{\apjl}{ApJ}           
\newcommand{\mnras}{MNRAS}       
\newcommand{\nat}{Nature}
\newcommand{\aap}{A\&A}
\newcommand{\araa}{ARA\&A}
\newcommand{\aj}{AJ}
\newcommand{\pasp}{PASP}
\newcommand{\apjs}{ApJS}           

\newcommand{\hi}{{\sc H\,i}}
\newcommand{\sauron}{\texttt{SAURON}}
\newcommand{\atl}{ATLAS$^{\rm 3D}$}
\newcommand{\kms}{\hbox{km s$^{-1}$}}
\newcommand{\msun}{\hbox{$M_\odot$}}
\newcommand{\lsun}{\hbox{$L_\odot$}}
\newcommand{\re}{\hbox{$R_{\rm e}$}}
\newcommand{\plotone}[1]{\includegraphics[width=\columnwidth]{#1}}
\newcommand{\refsec}[1]{Section~\ref{#1}}
\newcommand{\reffig}[1]{Fig.~\ref{#1}}

\title[The \atl\ project -- VII. The kinematic morphology-density relation]
{The \atl\ project -- VII. A new look at the morphology of nearby galaxies: the kinematic morphology-density relation}

\author[M.~Cappellari et al.]
{Michele Cappellari$^1$\thanks{E-mail: cappellari@astro.ox.ac.uk},
Eric Emsellem$^{2,3}$,
Davor Krajnovi\'c$^2$,
Richard M. McDermid $^{4}$,\newauthor
Paolo Serra$^{5}$,
Katherine Alatalo$^6$,
Leo Blitz$^6$,
Maxime Bois$^{2,3}$,
Fr\'ed\'eric Bournaud$^{7}$,\newauthor
M.~Bureau$^1$,
Roger L. Davies$^1$,
Timothy A. Davis$^1$,
P. T. de Zeeuw$^{2,8}$,\newauthor
Sadegh Khochfar$^{9}$,
Harald Kuntschner$^{10}$,
Pierre-Yves Lablanche$^{3}$,\newauthor
Raffaella Morganti$^{5,11}$,
Thorsten Naab$^{12}$,
Tom Oosterloo$^{5,11}$,
Marc Sarzi$^{13}$,\newauthor
Nicholas Scott$^1$,
Anne-Marie Weijmans$^{14}$\thanks{Dunlap Fellow}
and Lisa M. Young$^{15}$\\
$^1$Sub-department of Astrophysics, Department of Physics, University of Oxford, Denys Wilkinson Building, Keble Road, Oxford OX1 3RH\\
$^2$European Southern Observatory, Karl-Schwarzschild-Str. 2, 85748 Garching, Germany\\
$^3$Universit\'e Lyon 1, Observatoire de Lyon, Centre de Recherche Astrophysique de Lyon\\ and Ecole Normale Sup\'erieure de Lyon, 9 avenue Charles Andr\'e, F-69230 Saint-Genis Laval, France\\
$^{4}$Gemini Observatory, Northern Operations Centre, 670 N. A`ohoku Place, Hilo, HI 96720, USA\\
$^{5}$Netherlands Institute for Radio Astronomy (ASTRON), Postbus 2, 7990 AA Dwingeloo, The Netherlands\\
$^6$Department of Astronomy, Campbell Hall, University of California, Berkeley, CA 94720, USA\\
$^7$Laboratoire AIM Paris-Saclay, CEA/IRFU/SAp – CNRS – Universit\'e Paris Diderot, 91191 Gif-sur-Yvette Cedex, France\\
$^{8}$Sterrewacht Leiden, Leiden University, Postbus 9513, 2300 RA Leiden, the Netherlands\\
$^{9}$Max-Planck Institut f\"ur extraterrestrische Physik, PO Box 1312, D-85478 Garching, Germany\\
$^{10}$Space Telescope European Coordinating Facility, European Southern Observatory, Karl-Schwarzschild-Str. 2, 85748 Garching, Germany\\
$^{11}$Kapteyn Astronomical Institute, University of Groningen, Postbus 800, 9700 AV Groningen, The Netherlands\\
$^{12}$Max-Planck Institut f\"ur Astrophysik, Karl-Schwarzschild-Str. 1, 85741 Garching, Germany\\
$^{13}$Centre for Astrophysics Research, University of Hertfordshire, Hatfield, Herts AL1 9AB, UK\\
$^{14}$Dunlap Institute for Astronomy \& Astrophysics, University of Toronto, 50 St. George Street, Toronto, ON M5S 3H4, Canada\\
$^{15}$Physics Department, New Mexico Institute of Mining and Technology, Socorro, NM 87801, USA
}

\date{Accepted 2011 February 24. Received 2011 February 16; in original form 2010 November 30}

\pagerange{\pageref{firstpage}--\pageref{lastpage}} \pubyear{2011}

\begin{document}
\label{firstpage}
\maketitle

\clearpage
\begin{abstract}
In  Paper~I of this series we introduced a volume-limited {\em parent} sample of 871 galaxies from which we extracted the \atl\ sample of 260 ETGs. In Paper~II and III we classified the ETGs using their stellar kinematics, in a way that is nearly insensitive to the projection effects, and we separated them into fast and slow rotators. Here we look at galaxy morphology and note that the edge-on fast rotators generally are lenticular galaxies. They appear like spiral galaxies with the gas and dust removed, and in some cases are flat ellipticals (E5 or flatter) with disky isophotes. Fast rotators are often barred and span the same full range of bulge fractions as spiral galaxies. The slow rotators are rounder (E4 or rounder, except for counter-rotating disks) and are generally consistent with being genuine, namely spheroidal-like, elliptical galaxies. We propose a revision to the tuning-fork diagram by Hubble as it gives a misleading description of ETGs by ignoring the large variation in the bulge sizes of fast rotators.
Motivated by the fact that only one third (34\%) of the ellipticals in our sample are slow-rotators, we study for the first time the {\em kinematic} morphology-density $T-\Sigma$ relation using fast and slow rotators to replace lenticulars and ellipticals. We find that our relation is cleaner than using classic morphology. Slow rotators are nearly absent at the lowest density environments ($f(\rm SR)\la2\%$) and generally constitute a small fraction ($f(\rm SR)\approx4\%$) of the total galaxy population in the relatively low density environments explored by our survey, with the exception of the densest core of the Virgo cluster ($f(\rm SR)\approx20\%$). This contrasts with the classic studies that invariably find significant fractions of (misclassified) ellipticals down to the lowest environmental densities. We find a clean log-linear relation between the fraction $f(\rm Sp)$ of spiral galaxies and the local galaxy surface density $\Sigma_3$, within a cylinder enclosing the three nearest galaxies. This holds for nearly four orders of magnitude in the surface density down to $\Sigma_3\approx0.01$ Mpc$^{-2}$, with $f(\rm Sp)$ decreasing by 10\% per dex in $\Sigma_3$, while $f(\rm FR)$ correspondingly increases. The existence of a smooth  kinematic $T-\Sigma$ relation in the field excludes processes related to the cluster environment, like e.g.\ ram-pressure stripping, as main contributors to the apparent conversion of spirals into fast-rotators in low-density environments. It shows that the segregation is driven by local effects at the small-group scale. This is supported by the relation becoming shallower when using a surface density estimator $\Sigma_{10}$ with a cluster scale.
Only at the largest densities in the Virgo core does the $f(\rm Sp)$ relation break down and steepens sharply, while the fraction of slow-rotators starts to significantly increase. This suggests that a different mechanism is at work there, possibly related to the stripping of the gas from spirals by the hot intergalactic medium in the cluster core and the corresponding lack of cold accretion.
\end{abstract}


\section{Introduction}

Models of galaxy formation in the $\Lambda$CDM paradigm \citep{White1978} predict a strong dependence of galaxies properties on environment. As galaxies evolve via mergers one expects a faster galaxy evolution in groups or clusters, where the interactions or tidal disturbance are more frequent than in the isolated field and where the denser hot diffuse gas can strip galaxies of their cold gas \citep[e.g.][]{Blumenthal1984,Lacey1993,Kauffmann1993,Moore1996,
Kauffmann1999,Cole2000,Diaferio2001,Springel2005nat}.

Since the pioneering papers by \citet{Oemler1974} and \citet{Davis1976} discovering a dependence of galaxy morphology on environment and the classic work by \citet{Dressler1980}, which discovered the clear and nearly universal $T-\Sigma$ relation from a sample of about 6000 galaxies in 55 clusters, galaxy density has often been correlated to galaxy properties in large surveys to provide constraints on and test galaxy formation models \citep[see][for a review]{Blanton2009}. Subsequent observations for local galaxies generally confirmed the original finding of a tight and nearly universal $T-\Sigma$ relation in all environments, extending to the group environment \citep[e.g][]{Postman1984,Giovanelli1986}, with samples of galaxies reaching up to $10^5$ objects in the case of the Galaxy Zoo project \citep{Bamford2009,Skibba2009}. Studies have quantified how morphology separately varies as a function of mass, colours and galaxy shape at given environmental density \citep{vanderWel2008,Bamford2009,Skibba2009,vanderWel2010}.
However a long standing open questions is still whether the $T-\Sigma$ relation is driven by the cluster/group environment, and related to the distance from the cluster centre, or to the local galaxy density, as originally proposed  \citep{Whitmore1993,Goto2003,Bamford2009}.

A large set of papers have extended the observations of the $T-\Sigma$ relation to higher redshift using visual morphologies obtained from Hubble Space Telescope images to study its evolution: initial studies focused on lower redshifts $z\la0.5$ \citep[e.g.][]{Dressler1997,Fasano2000,Treu2003,Wilman2009} but are now routinely performed around $z\sim1$ \citep[e.g.][]{Stanford1998,vanDokkum2000,Smith2005,
Postman2005,Cooper2006,Capak2007,Poggianti2008}.
There is general consensus for an observed smaller fraction of S0s and a correspondingly larger one of spirals at larger redshift in high density environments, while no evolution is observed in the field. This is interpreted as due to the fact that the morphological segregation proceeds faster, and thus affects earlier, the denser cluster environments \citep{Dressler1997,Fasano2000,Postman2005,Smith2005}. Moreover little evolution was observed at the largest masses, for galaxies generally classified as ellipticals, suggesting that the most massive galaxies are already in place beyond $z\sim1$ \citep{Stanford1998,Postman2005,Tasca2009}, while most of the observed evolution consists of a transformation of spirals into S0s \citep{Smith2005,Moran2007}.

In this paper we study the $T-\Sigma$ relation for the volume-limited nearly mass-selected \atl\ {\em parent} sample of 871 nearby ($D<42$ Mpc) galaxies with $M_K<-21.5$ mag (stellar mass $M_\star\ga6\times10^9$ \msun) that we introduced in \citet[hereafter Paper~I]{Cappellari2010} and whose properties are summarized in Table~\ref{tab:parent_specs}.
Our survey, with its relatively modest sample size can shed little new light on the $T-\Sigma$ relation itself. However (for the first time) we have high-quality integral-field observations of the stellar kinematics (Paper~I) for all the ETGs in the sample. This exquisite level of detail allows us to properly classify the ETGs into genuinely spheroidal systems, the slow rotators, and lenticular-like galaxies, the fast rotators, in a robust way that is nearly insensitive to projection effects \citep{Cappellari2007,Emsellem2007}.

The classification utilizes the specific angular momentum parameter defined in \citet{Emsellem2007} as
\begin{equation}
\lambda_R\equiv\frac{\langle R |V| \rangle}{\langle R \sqrt{V^2+\sigma^2} \rangle} = \frac{\sum_{n=1}^{N} F_n\, R_n |V_n|}{\sum_{n=1}^{N} F_n\,R_n  \sqrt{V_n^2+\sigma_n^2}},
\label{eq:lambda}
\end{equation}
where $F_n$ is the flux contained inside the $n$-th Voronoi bin and $V_n$ and $\sigma_n$ the corresponding measured line-of-sight mean stellar velocity and velocity dispersion. We adopt here an improved criterion which is based on the classification of the velocity maps presented in \citet[hereafter Paper~II]{Krajnovic2010} and is introduced in \citet[hereafter Paper~III]{Emsellem2010}. The slow rotators are defined as those having $\lambda_R(\re)<0.31\sqrt{\varepsilon}$, where \re\ is the effective (half light) radius (Paper~I) and $\varepsilon$ is the ellipticity within 1\re.
The kinematic classification eliminates the well known high fractions of E/S0 misclassification that affect standard morphological classification using photometry alone. Our aim is to quantify the influence of the misclassification in the $T-\Sigma$ relation and its interpretation.

\begin{table}
\caption{Main characteristics of the \atl\ parent sample}
\centering
\begin{tabular}{rl}
\hline
Survey Volume: & ${\rm Vol}=1.16\times10^5$ Mpc$^3$ \\
Galaxy $K$-band luminosity: & $L>8.2\times10^9$ $L_{\odot,K}$ \\
Galaxy stellar mass: & $M_\star\ga6\times10^9$ \msun \\
Galaxy $B$-band total mag: & $M_B\la-18.0$ mag\\
Galaxy SDSS $r$-band total mag: & $M_r\la-18.9$ mag\\
Total number of galaxies: & $N_{\rm gal}=871$ \\
Spiral and irregular galaxies: & $N_{\rm Sp}=611$ (70\%) \\
S0 galaxies in \atl\ ($T>-3.5$): & $N_{\rm S0}=192$ (22\%) \\
E galaxies in \atl\ ($T\le-3.5$): & $N_{\rm E}=68$ (8\%) \\
\hline
\label{tab:parent_specs}
\end{tabular}
\end{table}

In this paper in Section~2 we give an overview of some of the limitations of the morphological classification by \citet{Hubble1936} and we show that with the help from the kinematic classification one can derive an improved description of the morphology of nearby galaxies. Motivated by our findings in Section~3 we derive the {\em kinematic} morphology-density relation for the galaxies in our sample and briefly discuss its interpretation in Section~4. A summary is given in Section~5.

\section{Morphology of nearby galaxies}
\label{sec:morph}

\subsection{Limitations of Hubble's tuning-fork diagram}

In the \citet{Hubble1936} tuning-fork diagram for the morphological classification of galaxies \citep[see][for a review of the history of the development]{Sandage2005}, the S0 class is a transition class between the spiral classes and the elliptical one \citep{deVaucouleurs1959,Sandage1961}. The bulge fraction in spiral galaxies increases from the later spiral types Sc towards the earlier types Sa, which are connected to the S0 class in the middle of the tuning-fork. This arrangement gives the impression that S0 galaxies should all have large bulge fractions, intermediate between Sa and elliptical galaxies.

This is in contrast with the observations by \citet{Spitzer1951} who first reported in the literature the S0 classification by Hubble and stated that ``the galaxies classed together as S0 by Hubble represent actually a series of forms paralleling the series of normal spirals, Sa, Sb, Sc. But the galaxies of the S0 series contain no obscuring matter and presumably are therefore unable to develop spiral structure.'' This was acknowledged by \citet{Sandage1970}, who noted that ``S0 galaxies have the appearance of spirals without arms''.  This point was strongly reinforced by \citet{vandenBergh1976}, who pointed out that ``Normal spirals, which exhibit a strong display of Population I, and S0 galaxies, which contain few if any young stars, form parallel sequences. These two sequences differ primarily in their total gas content and hence in the mean age of their stellar populations''.

\citet{vandenBergh1976} went on by proposing a new trident-like classification scheme in which he introduced the morphological types S0a, S0b and S0c, which are parallel to the Sa, Sb, and Sc sequence, and look like spirals with the dust removed. The proposed classification also introduced the class of Anemic Spirals: Aa, Ab and Ac, being the spirals with small amounts of gas, intermediate between normal spirals and S0s. The existence of ``passive spirals'' and the difficulty of robustly recognizing S0 and spiral galaxies, especially in clusters, was also recognized by \citet{Koopmann1998,Dressler1999,Poggianti1999}. The existence of these intermediate spiral systems, recently named ``red spirals'', is now being clearly recognized in large galaxy surveys using multi-colour CCD imaging \citep{Goto2003,Moran2006,Wolf2009,Masters2010}.

Although a number of authors agreed on the morphological similarity of S0 and spirals, and the existence of an intermediate class, the Hubble tuning-fork diagram is still in general use. This is perhaps due to the fact that the flattening and bulge fraction cannot be robustly measured in non-edge-on S0s, due to the difficulty of inferring the galaxies inclination and due to the mathematical degeneracy in the deprojection of the  surface brightness, when the galaxy is not edge-on \citep{rybicki87}. This makes the trident-like scheme of \citet{vandenBergh1976} more difficult to apply in practice than the tuning-fork of \citet{Hubble1936}. However we show in the next section that our observations of the \atl\ parent sample strongly support the findings by \citet{Spitzer1951} and \citet{vandenBergh1976}.

\subsection{Morphology of fast and slow rotators}
\label{sec:atlas3d_morphology}

\begin{figure*}
\includegraphics[width=0.9\textwidth]{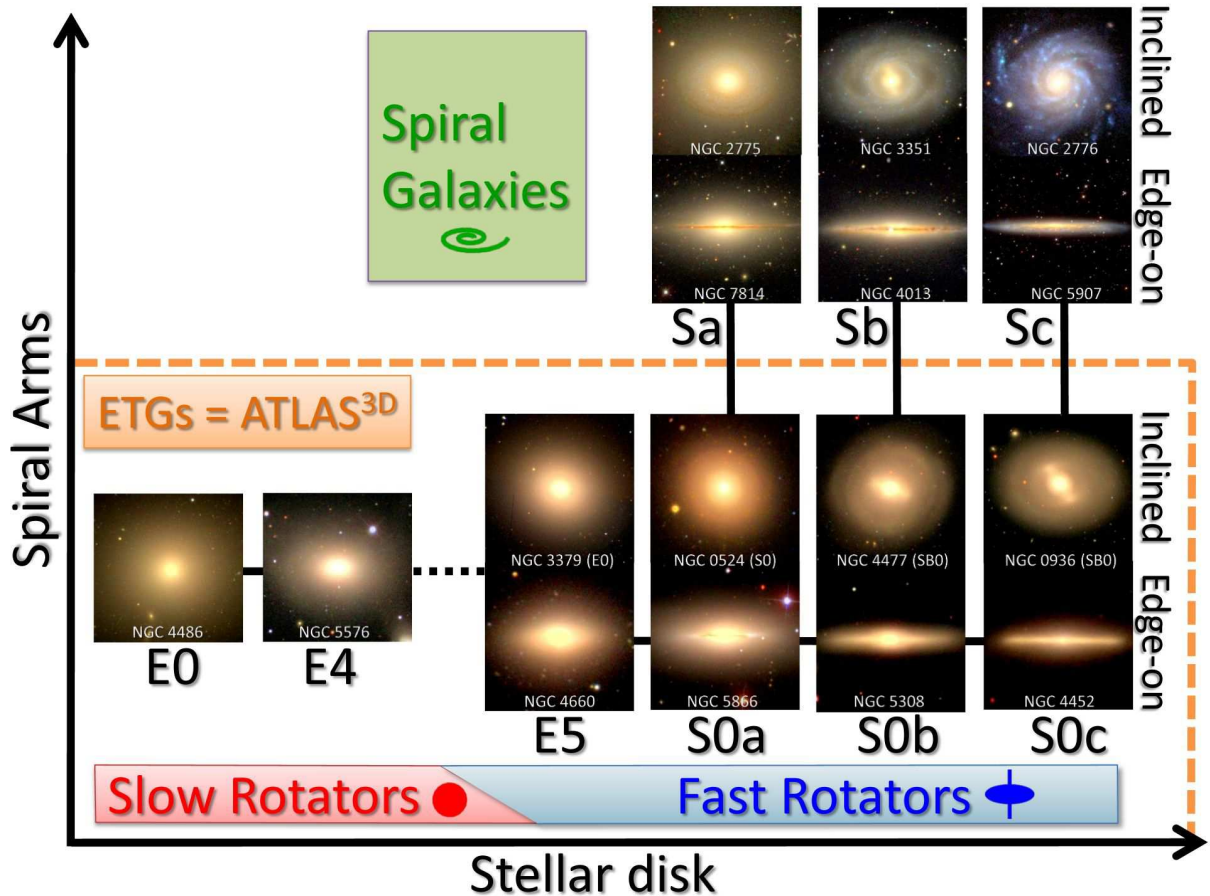}
\caption{Morphology of nearby galaxies from the \atl\ {\em parent} sample. The volume-limited sample consists of spiral galaxies (70\%), fast rotators ETGs (25\%) and slow rotators ETGs (5\%). The \atl\ sample consists of the ETGs only, classified according to the absence of spiral arms or an extended dust lane. The edge-on fast rotators appear morphologically equivalent to S0s, or to flat ellipticals with disky isophotes. Many of the apparently-round fast-rotators display bars or dusty disks, indicating that they are far from edge-on. All the galaxies classified as `disky' ellipticals E(d) by \citet{Bender94} belong to the fast-rotators class. However contrary to E(d) and S0 galaxies, the fast-rotators can be robustly recognized from integral-field kinematics even when they are nearly face-on \citep{Emsellem2007,Cappellari2007}. They form a parallel sequence to spiral galaxies as already emphasized for S0 galaxies by \citet{vandenBergh1976}, who proposed the above distinction into S0a--S0c. Fast rotators are intrinsically flatter than $\varepsilon\ga0.4$ and span the same full range of shapes as spiral galaxies, including very thin disks. However very few Sa have spheroids as large as those of E(d) galaxies.  The slow rotators are rounder than $\varepsilon\la0.4$, with the important exception of the flat S0 galaxy NGC~4550 (not shown), which contains two counter-rotating disks of nearly equal mass. The  black solid lines connecting the galaxy images indicate an empirical continuity, while the dashed one suggests a possible dichotomy.}
\label{fig:morphology_scheme}
\end{figure*}

\begin{figure*}
\includegraphics[width=0.7\textwidth]{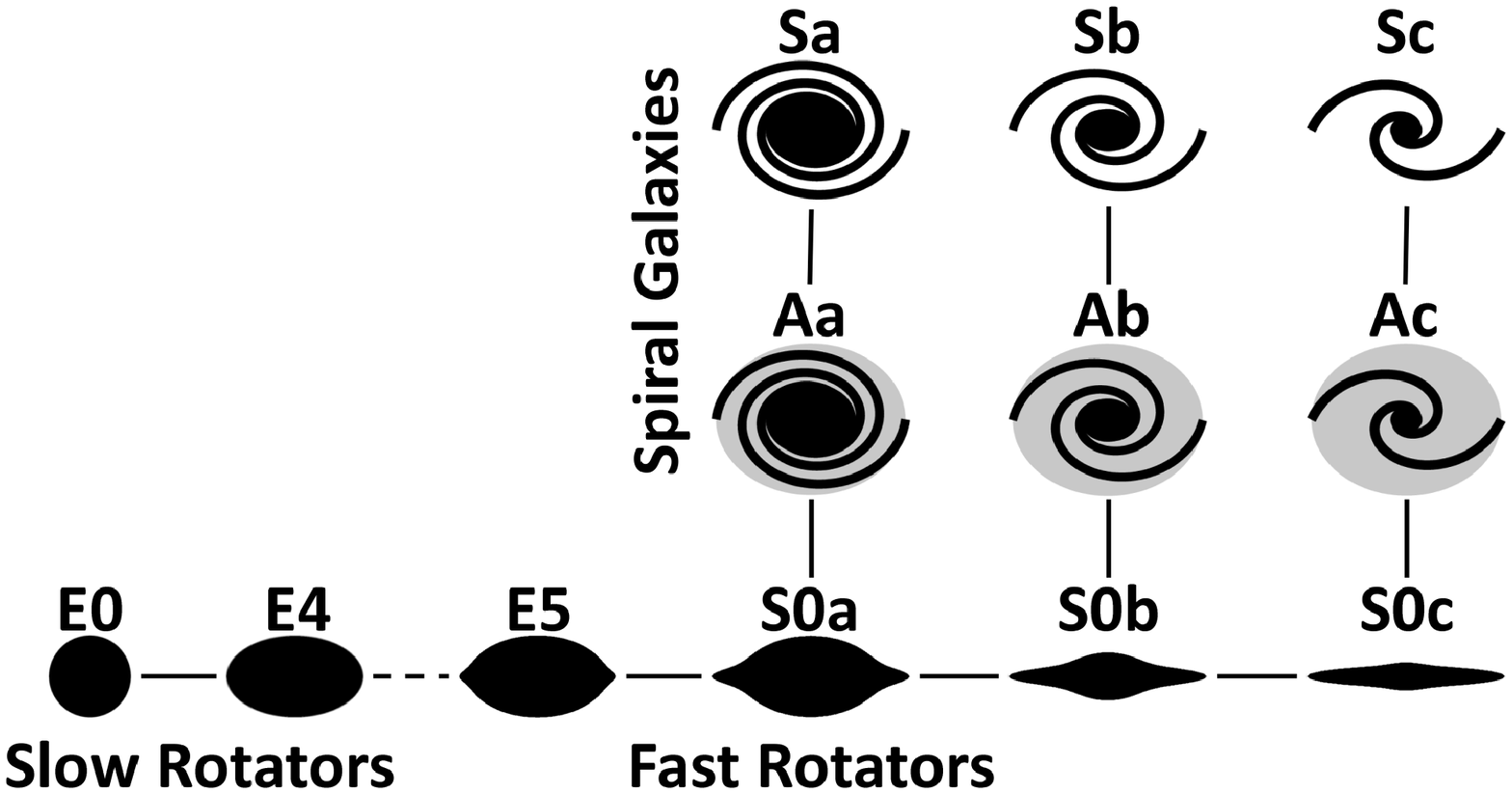}
\caption{Same as in \reffig{fig:morphology_scheme}, but in schematic form. As in previously proposed revisions \citep{vandenBergh1976,Kormendy1996} of Hubble's tuning-fork classification scheme, this diagram represents {\em intrinsic} galaxy properties. For this reason the slow-rotator (E0--E4) and fast-rotator (E5--S0c) early-type galaxies are visualized as edge-on.
Together with the spiral galaxies (Sa--Sc) and the early-type galaxies, here we also explicitly included in the diagram the class of Anemic Spirals (Aa--Ac) by \citet{vandenBergh1976}. These represent transition objects between the genuine spirals, with obvious large-scale spiral arms and the fast rotators, with no evidence of spiral structure in optical images.}
\label{fig:morphology_scheme_1}
\end{figure*}

A striking impression derived from inspection of the galaxies in the postage-stamp images of fig.~5 and 6 of Paper~I is the structural similarity between many nearly edge-on fast-rotator galaxies and spiral galaxies (not shown) in the parent sample. In particular for every fast rotator ETG that is known to be close to edge-on, from the presence of nuclear dusty disks, one can find a corresponding spiral galaxy with the same general shape, except for the presence of a prominent dust lane. This seems true even for the most spheroidal dominated fast rotators, although spirals with large spheroid are extremely rare in our sample. The morphological similarity is even more striking when looking at near-infrared $K$-band images from 2MASS \citep{Skrutskie2006}, where the dust absorption is essentially invisible. This is in agreement with the classification scheme proposed by \citet{Spitzer1951} and \citet{vandenBergh1976}, if one associates our fast-rotators with S0 galaxies. The large variation of bulge fractions of S0 and their overlap with the one of spirals was recently confirmed and accurately quantified by \citet{Laurikainen2010}. A number of anemic spiral galaxies with only weak evidence of dust is also present in the \atl\ {\em parent} sample, also in agreement with \citet{vandenBergh1976}.
The tuning-fork of \citet{Hubble1936} instead completely ignores the strong variations in the bulge fractions of S0 galaxies and does not allow for transition objects between the flat S0 and disk dominated Sc spirals. A modified version of \citet{vandenBergh1976} scheme, which illustrates the morphology of the parent sample is presented using postage-stamp images of real galaxies in \reffig{fig:morphology_scheme} and in schematic form in \reffig{fig:morphology_scheme_1}. Apart from morphology, other galaxy properties vary smoothly along this comb-shaped classification scheme. In fact, in an {\em average} sense, galaxy luminosity decreases from left to right and colours become bluer from the bottom to the top (e.g. figures~4 and 7 of Paper~I).

In addition to the connection between the morphology of fast-rotators and spiral galaxies, our diagrams also indicate an empirical continuity between the morphology of S0 fast-rotator galaxies, with increasing bulge fractions, and the flattest fast-rotator elliptical galaxies, which often show disky isophotes. The same concept was illustrated in the classification scheme of \citet{Kormendy1996}, where these galaxies are termed ``disky'' ellipticals E(d). All the galaxies classified as E(d) by \citet{Bender94} belong to the fast-rotator class. The complement however is not true as the weak disks of E(d) galaxies are only visible near the edge-on orientation, while the fast-rotator class can be recognized also near face-on view (Paper III).

The plot also illustrates the fact that the slow rotators appear to be intrinsically quite round (see fig.~5 of Paper I and Paper III), as already noticed in the \sauron\ survey \citep{Emsellem2007,Cappellari2007}. The only slow-rotator flatter than E4 in the \atl\ sample, and treated as `exception' in the diagram, is the S0 galaxy NGC~4550, which was indicated by \citet{Rubin1992} and \citet{Rix1992} for containing two counter-rotating disks of comparable mass. A detailed dynamical model of this galaxy, confirming the original interpretation and the nearly equal mass for the two disks was presented in \citet{Cappellari2007}. This object is not unique: a similar one (NGC~4473), classified as a fast-rotator due to the smaller fraction of counter-rotating stars, was also modeled by \citet{Cappellari2007} and a number of additional ones were newly discovered in \atl\ (Paper II), where they are termed ``double $\sigma$'' galaxies. Most of them are classified as fast-rotators, but some other are rounder slow rotators (Paper III). The resulting classification of this special class of objects seems to depend on the amount of accreted counter-rotating mass and the geometry of the orbit during the accretion event \citep[hereafter Paper~VI]{Bois2011}

The ellipticity distribution in the outer parts of the galaxies in our sample (Paper II) is characterized by a roughly constant fraction of galaxies up to ellipticities $\varepsilon\approx0.75$. Under the reasonable assumption of random orientations for the galaxies in our sample, this indicates that most of the galaxies, even when they appear round in projection, must possess quite flat disks as previously reported for S0 galaxies \citep{Sandage1970,Binney1981}. This is confirmed via Monte Carlo simulations in Paper III, while a quantitative statistical study of the shape of fast rotators will be presented in another paper of this series. This implies that the sample galaxies shown in \reffig{fig:morphology_scheme} are not exceptions, but are representative of our ETGs sample. Additional indications of the flatness of most of the galaxies in our sample comes from the fact that the inclination of the galaxies as derived from a sub-sample of objects with nuclear dusty disks agrees with the inclination derived from the shape of the outer isophotes, under the assumption of a flat disk \citep[hereafter Paper~V]{Davis2010}.

A recent dynamical modeling study suggests that the stellar dynamics of the fast rotators appear to be indistinguishable from that of spiral galaxies of comparable flattening \citep{Williams2009}, both being well described by the simple anisotropic Jeans models of \citet{Cappellari08}. Within the limits of the small size of the studied sample this further confirms the structural similarity of the two classes of objects. From a purely empirical point of view the difference between fast rotator ETGs and spiral galaxies is in their dust content, visible as spiral arms in optical photometry, and their cold gas content, as detected via molecular lines, which is significantly lower than spiral galaxies \citep[hereafter Paper~IV]{Young2010}.

The similarity of the morphologies illustrated in \reffig{fig:morphology_scheme} and \ref{fig:morphology_scheme_1} however does not implies a similarity in the distribution of bulge fractions, which appears qualitatively different between spirals and fast-rotators, with the latter being on average characterized by larger bulges. This was quantified for S0 galaxies by \citet{Simien1986} and recently by \citet{Laurikainen2010}, who note that S0 tend to have bulge fractions comparable to that of Sa galaxies, but larger than later spiral types. The same is expected to hold for fast rotators, which contain large fractions of S0 galaxies.
An indication of this trend for the galaxies of our sample was illustrated in fig.~4 of Paper I, which shows the $K$-band size-luminosity relation for the parent sample as a function of the morphological type.
The fast-rotators indeed have a significant overlap in this diagram with the spiral galaxies of type Sa--Sb. However there is a clear trend in the $\re-L_K$ diagram as a function of galaxy morphology. The observed trend is due to a variation in the bulge size, with bulges progressively increasing (by definition) from Sd-Sc to Sb-Sa and to fast rotators ETGs.

The novelty of our project, with respect to all previous morphological studies, is that it provides integral-field observations of the stellar kinematics for all the galaxies. This allows us to recognize disk-like systems even when they are close to face on and could be morphologically misclassified as genuine elliptical galaxies. The close connection illustrated in \reffig{fig:morphology_scheme} and \ref{fig:morphology_scheme_1} between (i) slow-rotators and genuine elliptical galaxies and (ii) fast-rotators and S0-like galaxies, all seen at various orientations, is the motivation for revisiting the classic morphology-density $T-\Sigma$ relation in the next section.

\section{The morphology-density relation}

\subsection{Three estimators of galaxies environment}
\label{sec:environment}

\begin{figure}
\plotone{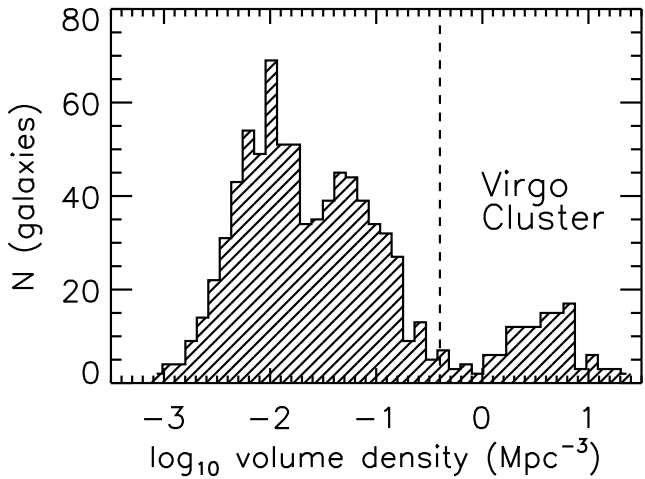}
\caption{Histogram of the local density, as measured by $\rho_{10}$, for all the galaxies of the \atl\ parent sample. There is a clear bimodality of the distribution, above or below $\log\rho_{10}\approx-0.4$. Nearly all galaxies in Virgo have $\log\rho_{10}>-0.4$, while the ones outside Virgo have $\log\rho_{10}<-0.4$.}
\label{fig:density_histogram}
\end{figure}

\begin{figure}
\plotone{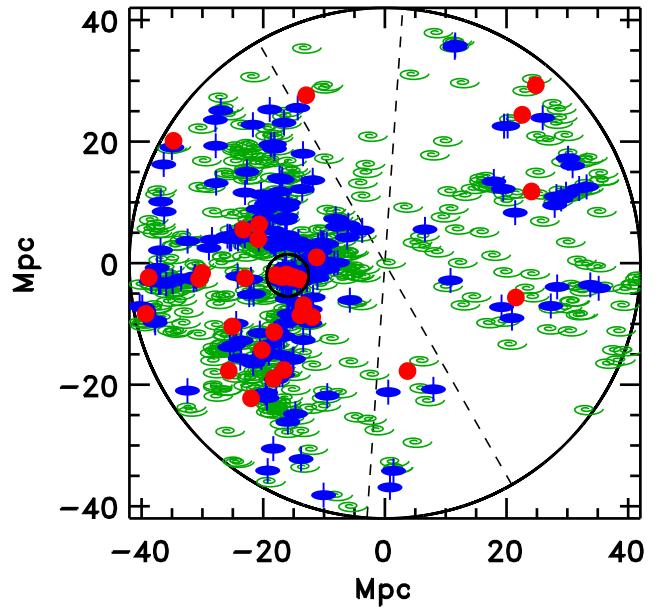}
\caption{Projection of the local volume enclosing the \atl\ parent sample. The blue ellipses with axis are fast-rotators ETGs, the red filled circles are slow rotators ETGs, while the green spirals are spiral galaxies. The Earth equatorial plane lies on the $x-y$ plane, with the $x$-axis points towards ${\rm RA}=0$. The small black circle centered on Virgo has a radius $R=3.5$ Mpc, while the large one defines our selection volume with $R=42$ Mpc. The dashed lines indicate the intersection between the (inclined) Galaxy zone-of-avoidance and the $x-y$ plane. This exclusion zone explains the most empty region of the plot. The filamentary nature of galaxy clustering and the large range of volume  densities sampled by the survey are evident.}
\label{fig:volume_projection}
\end{figure}

For all galaxies of the \atl\ parent sample we computed a set of density parameters describing the number and luminosity of galaxies around each galaxy. For our density estimate we followed the same steps as in Paper~I to extract the galaxy sample and derive distances. However here we considered all galaxies in a volume which fully encloses that defined by the parent sample with $M_K<-21.5$ mag and $D<56$ Mpc ($V_{\rm hel}<4000$ \kms), ignoring observability selections, to prevent possible incompleteness at the edges of our volume. As recession velocities for the ETGs we used our own accurate \sauron\ heliocentric velocities $V_{\rm hel}$ from stellar kinematics (table~3 of Paper~I), instead of the NED values.

We defined two types of density estimators. They are both based on the adaptive method of \citet{Dressler1980}, but one is measured inside spheres, while the others are determined inside redshift cylinders. In all cases the density refers to the subset of galaxies with $M_K<-21.5$ mag, which is the brightness limit of the survey. All these quantities are tabulated in Table~\ref{tab:atlas3d_sample} and Table~\ref{tab:atlas3d_spirals} for the ETGs and spiral galaxies in the parent sample:
\begin{enumerate}
\item $\rho_{10}=N_{\rm gal} / (\frac{4}{3}\pi r_{10}^3)$ is the volume density in Mpc$^{-3}$ of galaxies inside a sphere of radius $r_{10}$ centred on a galaxy, which includes $N_{\rm gal}=10$ nearest neighbours. For all estimators $N_{\rm gal}$ excludes the galaxy under consideration. For this estimator we adopted the best distance estimates and the sky coordinates of all the galaxies in the parent sample to compute the three-dimensional Cartesian coordinates of all the galaxies inside the local volume. This allows for an accurate density estimation in Virgo, where a number of accurate distances, based on SBF are available \citep{Mei2007}. For the Virgo galaxies without SBF distance, to avoid overestimating the local density, this was estimated after adding a Gaussian scatter with $\sigma=0.6$ Mpc, as measured by \citet{Mei2007}, to the Virgo distance of $D=16.5$ Mpc. The resulting median value of the sampled radius for our sample $r_{10}=4.5$ Mpc.
    The $\rho_{10}$ estimator has the limitation that a galaxy with inaccurate redshift-independent distance, inside a group of galaxies with only redshift distance, may incorrectly appear outside the group. In that case the density of galaxies around it will be underestimated. In practice we found this to be a problem only for a handful of galaxies.
\item $\Sigma_{10}=N_{\rm gal} / (\pi R_{10}^2)$ is the surface density in Mpc$^{-2}$ of galaxies inside a cylinder of radius $R_{10}$ and height $h=600$ \kms\ (i.e. $\Delta V_{\rm hel}<300$ \kms) centred on the galaxy, which includes $N_{\rm gal}=10$ nearest neighbours. Here we used the heliocentric velocity while ignoring redshift-independent distances. This estimator is essentially the same as defined by \citet{Dressler1980}. The resulting median value of the sampled radius for our sample $R_{10}=3.8$ Mpc.
\item $\Sigma_3=N_{\rm gal} / (\pi R_{3}^2)$ is the same as $\Sigma_{10}$, but considers a cylinder containing the $N_{\rm gal}=3$ nearest neighbours. The resulting median value of the sampled radius for our sample $R_{3}=1.3$ Mpc.
\end{enumerate}

In \reffig{fig:density_histogram} we show a histogram of the density $\rho_{10}$ for the full parent sample.
There is a clear bimodality, with a minimum around $\log\rho_{10}\approx-0.4$ and the majority of the galaxies below that value. It turns out that with only two exceptions, all galaxies in Virgo have $\log\rho_{10}>-0.4$, while those outside Virgo have $\log\rho_{10}<-0.4$. According to this estimator the sample includes differences of four orders of magnitudes between the high density (Virgo) and low density environments. The Virgo membership is given in table~3 and 4 of Paper~I and was defined as being within a radius $R=3.5$ Mpc from the centre of the cluster assumed at coordinates RA$=$12h28m19s and DEC$=$+12$^\circ$40$'$ \citep{Mould2000} and distance $D=16.5$ Mpc \citep{Mei2007}. The radius corresponds to an angular size of 12$^\circ$ on the sky.

\begin{figure}
\plotone{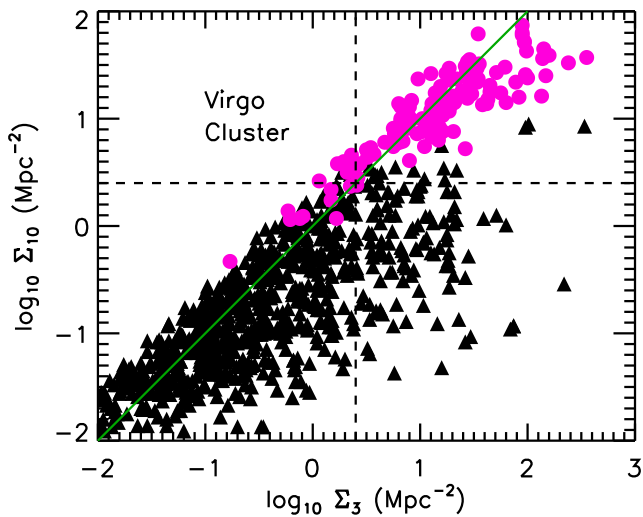}
\caption{Comparison between the density estimators $\Sigma_3$ and $\Sigma_{10}$. Virgo galaxies are indicated by the magenta filled circles, while non-virgo ones are black filled triangles. Using the $\Sigma_{10}$ estimator a rough separation between cluster/field is obtained at $\Sigma_{10}\approx0.4$, however no clean separation can be obtained using the $\Sigma_3$ estimator: for many bins of $\Sigma_3$ there are galaxies both inside and outside the cluster.}
\label{fig:morphology_density_10_3}
\end{figure}

A comparison between the $\Sigma_{10}$ surface density estimator and the more local $\Sigma_{3}$ is shown in \reffig{fig:morphology_density_10_3}. The plot illustrates the fact that, while the $\Sigma_{10}$ broadly separates the Virgo cluster environment from the rest, the $\Sigma_{3}$ estimator eliminates the distinction. Using $\Sigma_{3}$ one finds that some small groups of $N_{\rm gal}=4$ galaxies are as closely packed as galaxies in the core of Virgo, while many small groups are as dense as the outskirts of the Cluster.

To put our survey in a general context, we compare the density of galaxies in the survey, with the density of galaxies in other environments in the nearby universe. Using the numbers from Table~7.1 of \citet{Sparke2007} we find that inside the core of the Virgo cluster there is a density of about 560 galaxies Mpc$^{-3}$ above a luminosity $L_B=10^{9} \lsun$. This density is comparable to the one of 900 galaxies Mpc$^{-3}$ inside the core of the Fornax cluster but it is still significantly lower than the value of about 13,400 galaxies Mpc$^{-3}$ in the core of the Coma cluster.

A global view of the local galaxy volume of the \atl\ sample is given in \reffig{fig:volume_projection}. It shows the quite inhomogeneous distribution of ETGs in the local volume and the variety of environment sampled by the survey. In some cases galaxies appear to be associated to filaments, but the survey also includes extremely isolated objects. When the size of our volume is compared with the scale of the filamentary structure of dark matter predicted by numerical simulations \citep[e.g.][]{Springel2005nat}, one can see that our volume is expected to sample a large number of sub clusters and filaments, although the volume is still affected by cosmic variance (\citealt{Khochfar2011}, hereafter Paper~VIII).

\subsection{Specific angular momentum of ETGs versus local density}

\begin{figure}
\plotone{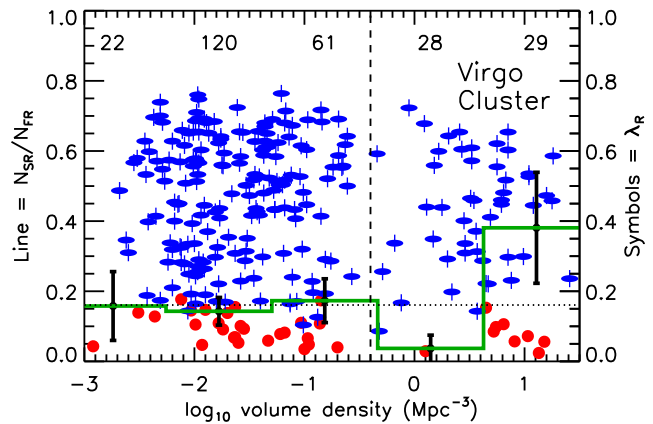}
\caption{$\lambda_R$ versus density. The specific angular momentum for all the galaxies in the \atl\ sample is plotted against the density $\rho_{10}$. The blue ellipses with vertical axis and the red circles represent the fast and slow rotators respectively (Paper III). The dashed vertical line marks the separation between galaxies inside ($\log\rho_{10}<-0.4$) and outside the Virgo Cluster. The green solid line with error bars indicates the ratio $N_{\rm SR}/N_{\rm FR}$ of slow rotator versus fast rotator galaxies as a function of $\rho_{10}$, while the dotted horizontal line is the global ratio $N_{\rm SR}/N_{\rm FR}$ for the survey. The fraction is nearly constant outside Virgo, but it shows a dramatic decrease in the outskirts of Virgo and an increase inside the cluster core.}
\label{fig:lambda_density}
\end{figure}

\begin{figure}
\plotone{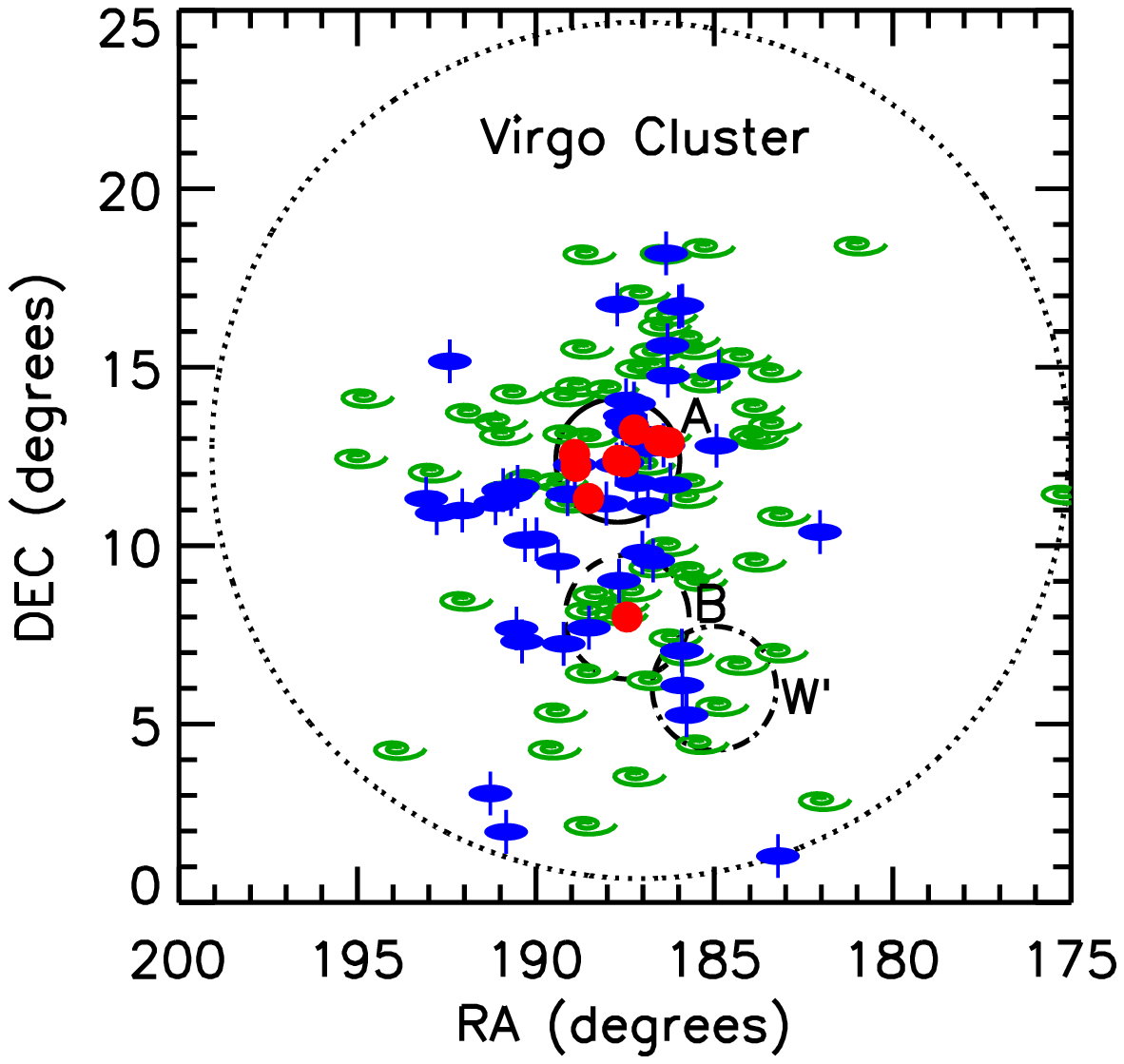}
\plotone{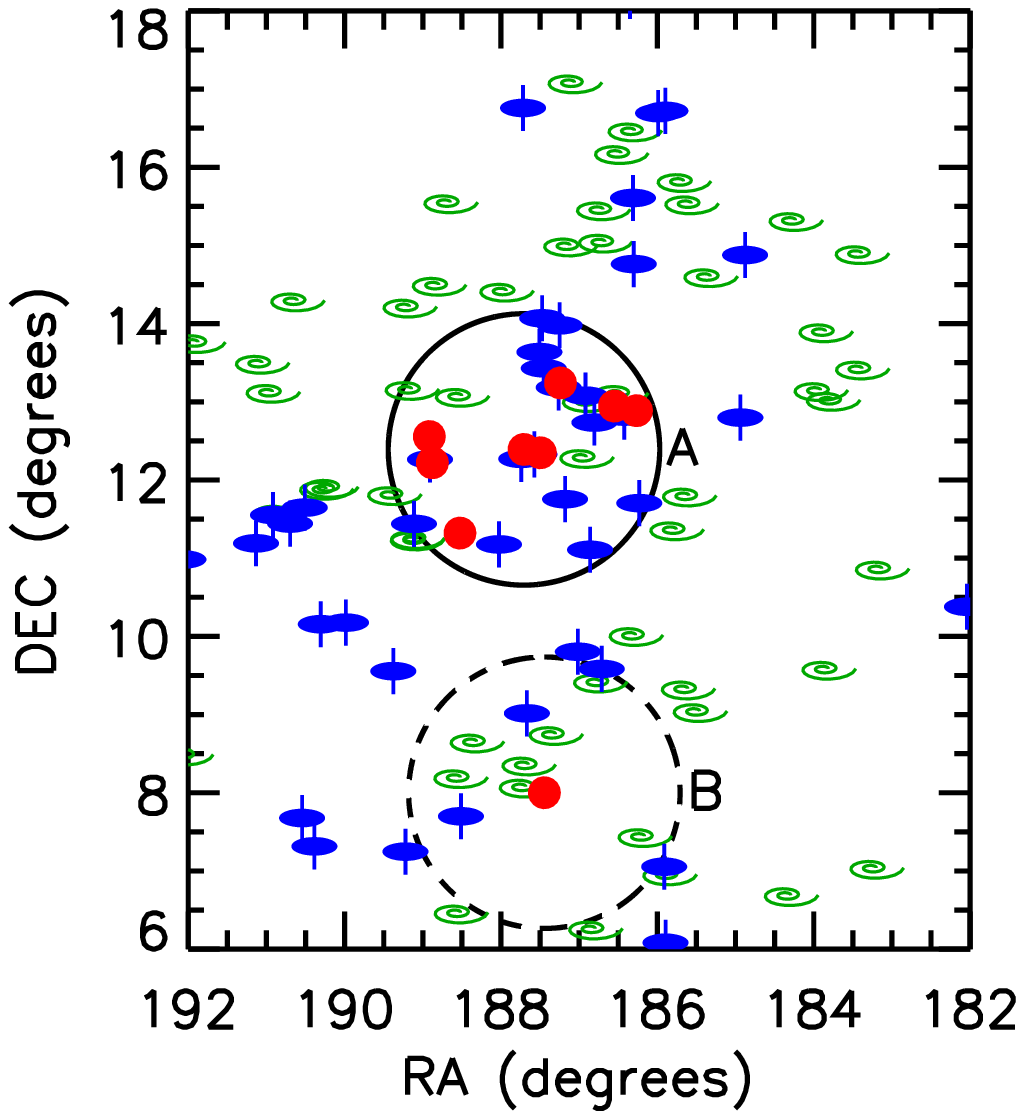}
\caption{Fast/slow rotators and spirals in the Virgo Cluster. The symbols are as in \reffig{fig:volume_projection}. All Virgo galaxies with $M_K<-21.5$ mag are shown. The large dotted circle centred on the Virgo core has a radius of 12$^\circ$ ($R=3.5$ Mpc) and defines our adopted limits of the cluster, the solid circle indicates cluster A, centred on M87, the dashed circle indicates cluster B, centred on M49, and the dash-dotted circle is the cloud W$'$ as defined in \citet{Binggeli1987}. The top panel shows the full cluster extent, while the bottom one zooms on the Virgo core. Out of 9 slow rotators in Virgo, 8 are concentrated inside the core of cluster A (solid circle with $R=0.5$ Mpc) and one is M49, which defines the centre of cluster B. The fast rotators are more uniformly distributed. The spiral galaxies overlap with the fast rotators, but tend to be more common in the outskirts of the cluster.}
\label{fig:lambda_virgo}
\end{figure}

Here we revisit the $T-\Sigma$ relation that was studied in the past, using the kinematical classification based on $\lambda_R$, instead of the classic morphology. We first focus on the ETGs alone and we use the fast and slow rotators to replace S0 and elliptical galaxies, due to their closely related morphologies (\refsec{sec:atlas3d_morphology}). The fast/slow rotator classes and the $\lambda_R$ values for the galaxies in the \atl\ sample are given in Paper~III.

Here we study the dependence of $\lambda_R$ on the galaxy environment defined in \refsec{sec:environment}. In \reffig{fig:lambda_density} we plot $\lambda_R$ versus the density $\rho_{10}$.
We find that slow rotator galaxies are not only found in dense environments. If we restrict our analysis to the ETGs alone, the fraction of slow versus fast rotators is nearly insensitive to environment, outside the Virgo cluster, over more than two orders of magnitude in the density $\rho_{10}$, and within the errors is consistent with the global value $N_{\rm SR}/N_{\rm FR}\approx16\pm3\%$ found for the whole \atl\ sample. However inside the Virgo cluster the situation changes dramatically and the slow rotators there are all segregated in the most dense environment, in the cluster core, where the fraction becomes $N_{\rm SR}/N_{\rm FR}\approx38\pm16\%$ compared to $N_{\rm SR}/N_{\rm FR}\approx4\pm4\%$ in the outer parts of the cluster.
The prevalence of slow rotators in the core of the Virgo cluster is already evident in \reffig{fig:lambda_virgo} which show the distribution on the sky of fast and slow rotators ETGs and spiral galaxies. A clear feature is that 8/9 of the slow rotators ETGs are contained within the innermost $R<0.5$ Mpc (small solid circle) from the cluster centre, while the fast rotators and spiral galaxies appear to share a similar and more uniform distribution.

The findings of this section suggest the presence of at least two different processes for the formation of slow rotators: (i) in the field or in small groups the slow rotators form via a quite inefficient process, which is nearly insensitive to the environment, (ii) while in dense cluster environment a much more efficient process is at work, which depends sensitively on the local density, or on the distance from the cluster core.

\subsection{The kinematic Morphology-Density relation}

In the previous section we saw that the ratio between fast and slow rotators has a very small sensitivity to environment, at least within the density ranges explored by our local volume, except in the Virgo core. One may wonder whether this is due to the fact that we do not explore a sufficiently wide range of densities. We show in this section that this is not the case and in fact the picture changes dramatically when one includes spiral galaxies in the study.

\begin{figure}
\includegraphics[width=0.95\columnwidth]{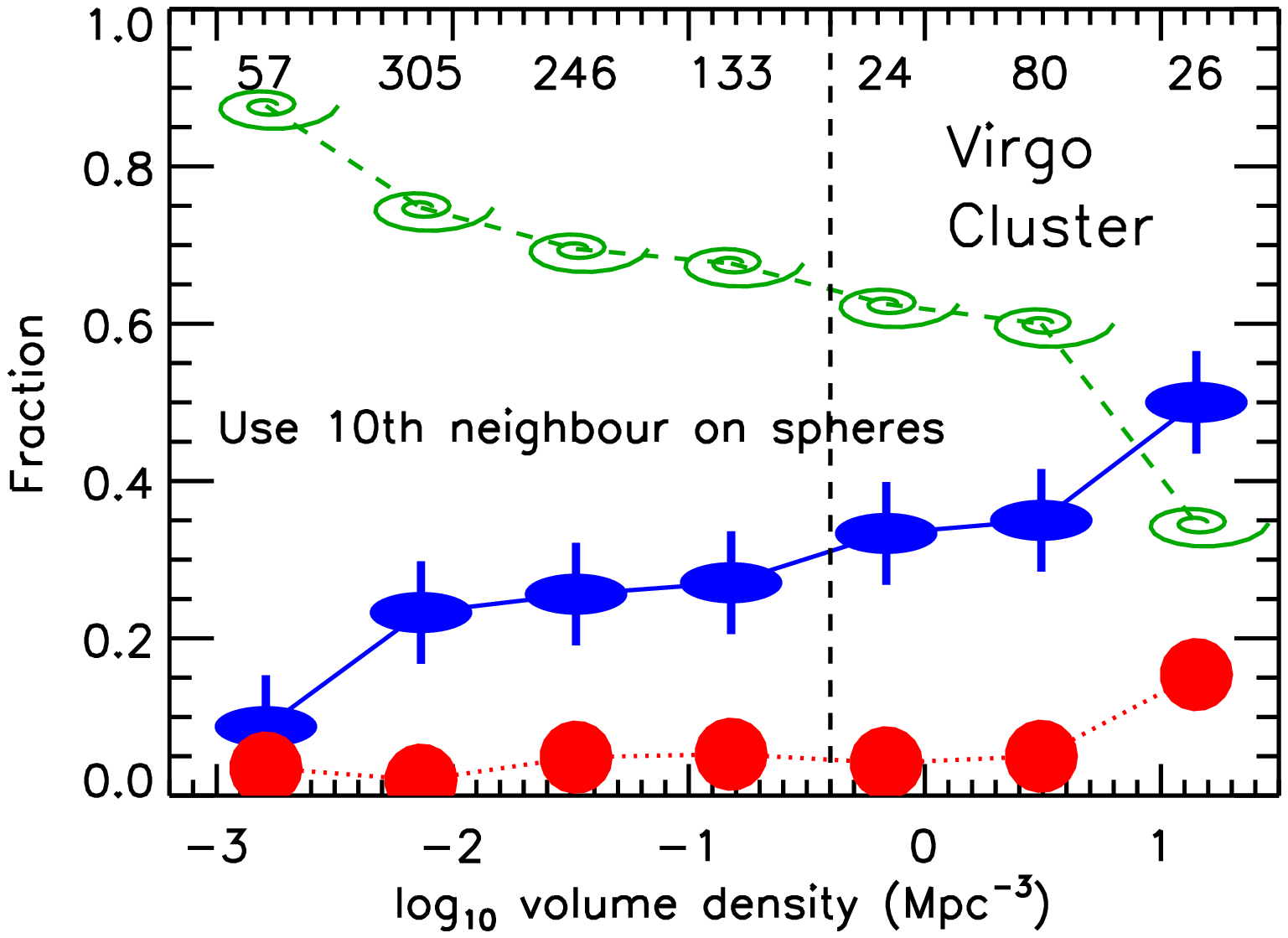}
\includegraphics[width=0.95\columnwidth]{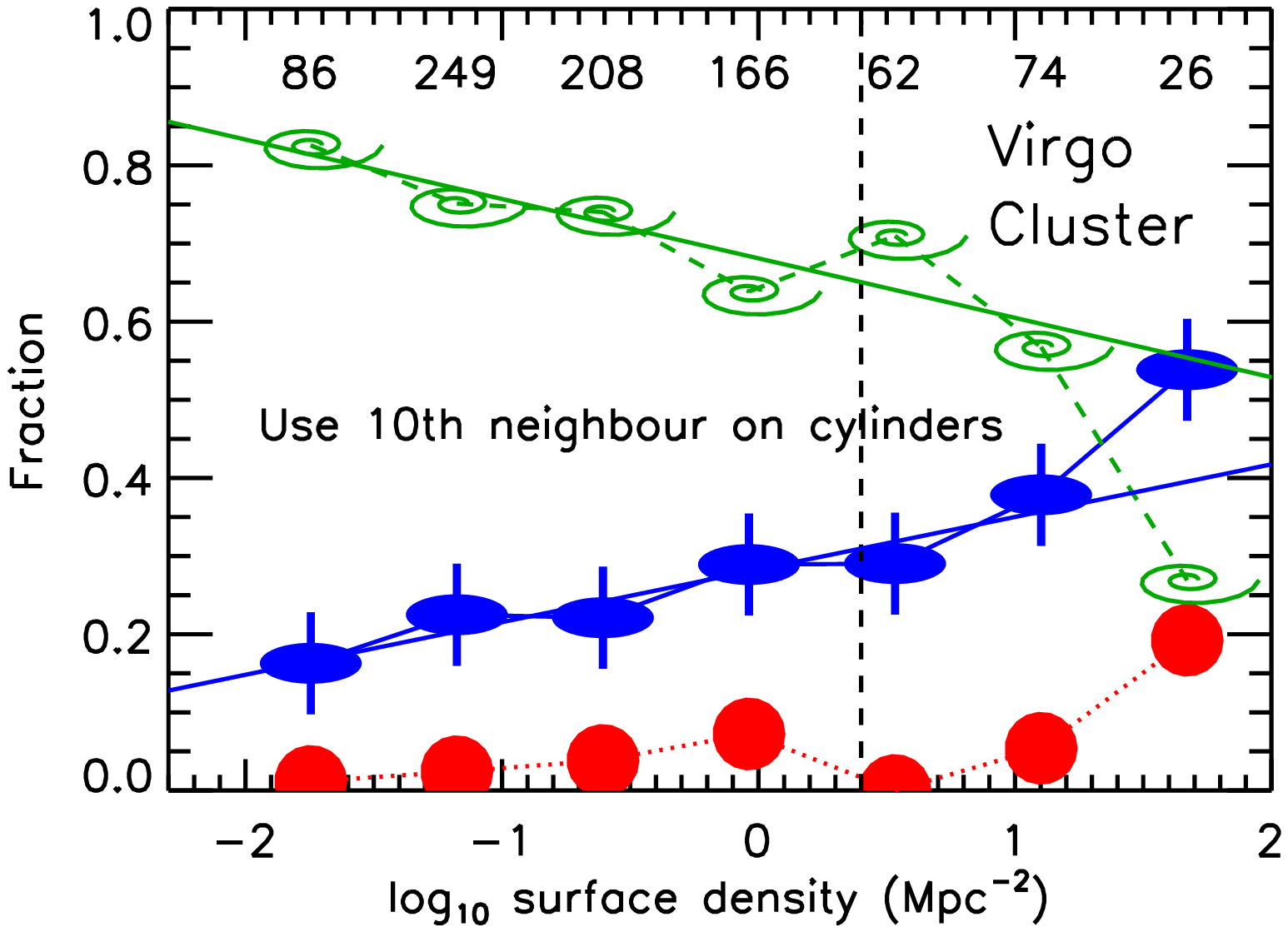}
\includegraphics[width=0.95\columnwidth]{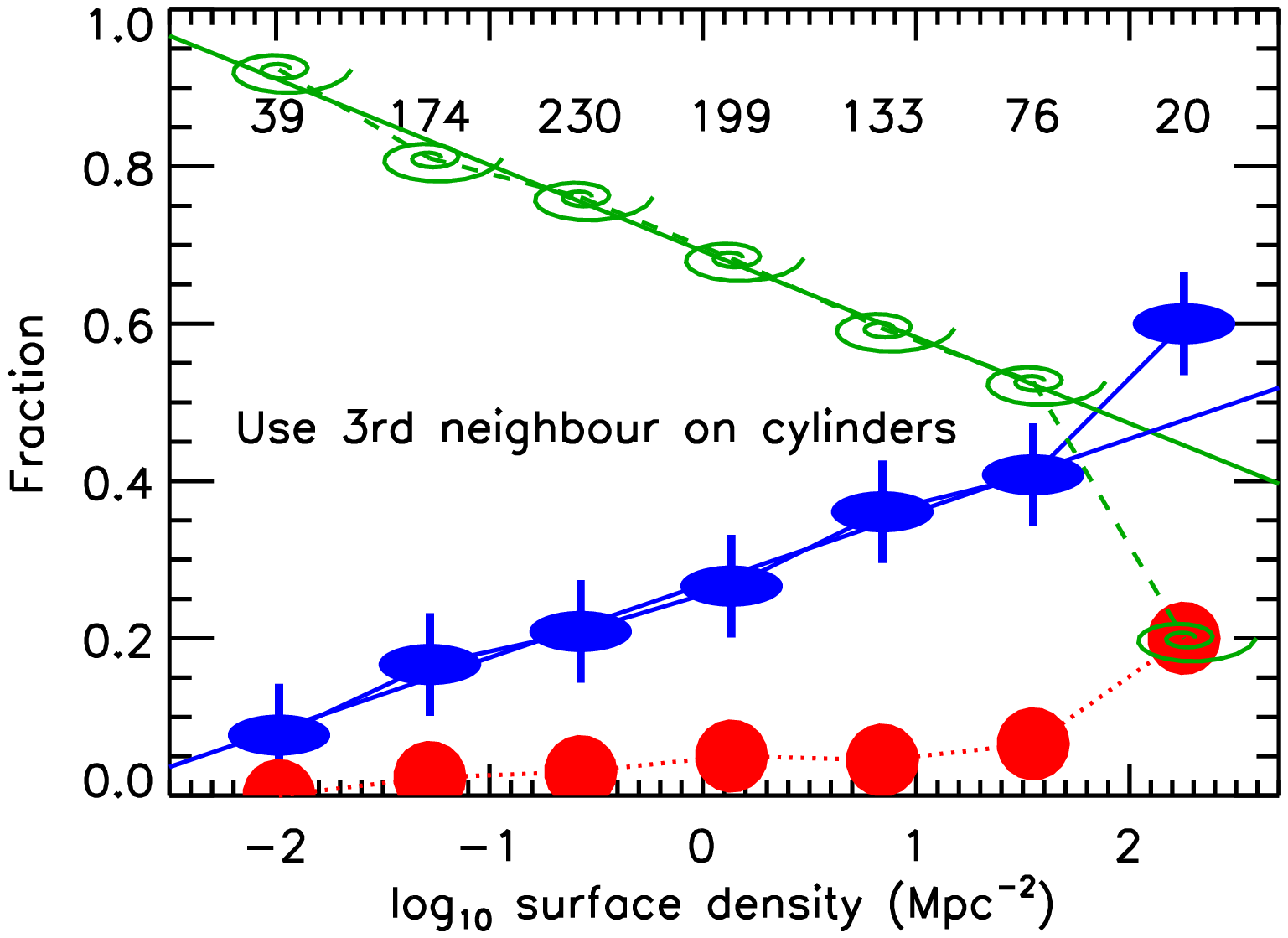}
\caption{The $T-\Sigma$ relation for fast rotators (blue ellipse with vertical axis), slow rotators (red filled circle) and spiral galaxies (green spiral). The dashed vertical line in the top two panels indicates an approximate separation between the density of galaxies inside/outside Virgo. In the bottom two panels the solid blue and magenta lines are best fit to the first six values. The numbers above the symbols represent the number of galaxies included in each of the seven density bins.}
\label{fig:morphology_density}
\end{figure}

In \reffig{fig:morphology_density} we show the $T-\Sigma$ relation for fast and slow rotator ETGs and spiral galaxies using, from top to bottom, the three different density estimators $\rho_{10}$, $\Sigma_{10}$ and $\Sigma_3$ respectively (\refsec{sec:environment}). In agreement with the previously reported studies we find a clear trend for the spiral fraction $f(\rm Sp)$ to gradually decrease with environmental density while the fraction of ETGs correspondingly increases. This trend continues smoothly over nearly four orders of magnitude in density and does not flatten out even at the lowest densities. The fraction of spirals is equal to that of ETGs at a volume density that corresponds to a region within the core of the Virgo cluster.

Broadly speaking the three density estimators provide qualitatively similar trends and we first focus on $\Sigma_3$, which provides the cleanest relation. We find two new results:
\begin{enumerate}
\item The extreme low densities explored by our local volume allow us to demonstrate that the most isolated galaxies are almost invariably spirals. In fact as much as $f(\rm Sp)=36/39=92\%$ of the galaxies in the lowest density bin ($\Sigma_{3}=0.01$ Mpc$^{-2}$) are spirals. Considering the two lowest $\log\Sigma_3$ density bins, to improve the statistics, we find $f(\rm Sp)=177/213=83\%$. This spiral fraction is consistent with the estimate for the AMIGA sample of isolated galaxies \citep{Sulentic2006};
\item The use of our kinematic classification shows that genuine spheroidal ETGs, the slow-rotators, make up only a very small fraction ($f(\rm SR)\approx4\%$) of the total galaxy population except in the Virgo core where they contribute to $\approx20\%$ of the total. The slow rotators contribute even lower fractions at the lowest densities: there are no slow rotators in the lowest density bin, while considering the two lowest $\log\Sigma_3$ bins we find $f(\rm SR)=4/213=1.9\%$. Of the four slow rotators in the two bins, one (NGC~6703) is indicated in Paper~III as a possible face-on fast rotators and another one (UGC~03960) has a low data quality. This implies that the fraction of genuine slow rotators in the two lowest density bins may be as low as 1\%. This is in strong contrast to traditional studies of the $T-\Sigma$ relation that never find less than $\sim10\%$ of (misclassified) elliptical galaxies even in the lowest density environments \citep[e.g.][]{Postman1984,Bamford2009}.
\end{enumerate}

Looking in more detail, there is a notable difference between the $T-\Sigma$ relation obtained using the $\Sigma_{10}$ and $\Sigma_3$ estimators.
Using both estimators the fraction of spirals $f(\rm Sp)$ and fast-rotators $f(\rm FR)$ are well described by two linear relations of $\log\Sigma$ (see also \citealt{Dressler1997}). However using $\Sigma_3$ the relations become noticeably steeper and more cleanly defined. Moreover using $\Sigma_3$ the fraction of slow rotators $f(\rm SR)$ does not show the drop that is observed using $\Sigma_{10}$ in the outskirts of the Virgo cluster (as pointed out regarding \reffig{fig:lambda_virgo}). The best fitting $T-\Sigma$ relations using the two surface-density estimators are:
\begin{equation}
    f(\rm Sp) = 0.69 - 0.07\times\log\Sigma_{10}
\end{equation}
\begin{equation}
    f(\rm FR) = 0.28 + 0.06\times\log\Sigma_{10}.
\end{equation}
using the $\Sigma_{10}$ estimator and restricting the linear fit to the range $0.01\la\Sigma_{10}\la20$ Mpc$^{-2}$, and
\begin{equation}
    f(\rm Sp) = 0.69 - 0.11\times\log\Sigma_{3}
\end{equation}
\begin{equation}
    f(\rm FR) = 0.26 + 0.09\times\log\Sigma_{3}.
\end{equation}
using the $\Sigma_3$ estimator in the range $0.01\la\Sigma_{3}\la50$ Mpc$^{-2}$. The relations for $f(\rm Sp)$ and $f(\rm FR)$ have nearly opposite slopes so that the decrease of spirals is compensated by an increase of fast-rotators, with an insignificant contribution from slow rotators.

\begin{figure}
\plotone{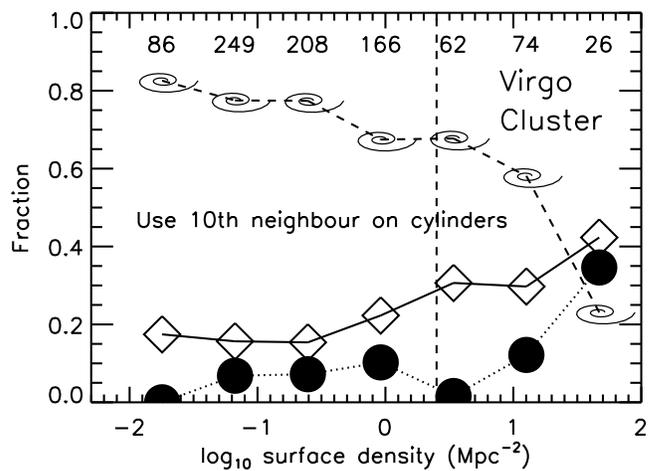}
\caption{Morphology versus density for elliptical (black filled circles), lenticular (open diamonds) and spiral galaxies (spirals), versus the local surface density $\Sigma_{10}$. The numbers above the symbols represent the number of galaxies included in each of the seven density bins.}
\label{fig:morphology_density_classic}
\end{figure}

To provide a direct connection of our work with the ones of previous authors on the $T-\Sigma$ relation, in \reffig{fig:morphology_density_classic} we show the usual relation using the morphological classification given by the HyperLeda\footnote{http://leda.univ-lyon1.fr/} database \citep{Paturel2003}, using the morphological $T$-type (tabulated in Paper~I) where ellipticals have $T\le-3.5$, S0s have $-3.5<T\le-0.5$ and spirals have $-0.5<T$. The adopted surface density estimator $\Sigma_{10}$ closely reproduces the one introduced by \citet{Dressler1980}.
The spiral fraction is nearly identical to the one in the middle panel of \reffig{fig:morphology_density}, as expected, given the excellent agreement between our morphology and the HyperLeda one. However the ratio between the ellipticals and S0 appears to be quite different from that between slow-rotators and fast-rotators. Of the 68/871 elliptical galaxies in our volume-limited sample, only 23/871 are genuinely spheroidal slow rotators (even fewer if one excludes the counter-rotating disks like NGC~4550). This implies that only about one third (34\%) of the elliptical galaxies are properly classified using photometry alone, while the majority of them is composed of misclassified disk-like fast-rotators. This is qualitatively consistent with the statistical estimations of elliptical misclassifications made twenty years ago \citep{Rix1990,vandenBergh1990,Jorgensen1994,Michard1994}. However the key difference of our kinematic classification is that we can finally properly classify ETGs on an individual basis.

\section{Discussion}

One possible explanation for the existence of a smoothly decreasing $T-\Sigma$ relation for $f(\rm Sp)$, over nearly four orders of magnitudes in $\Sigma_3$, matched by the increase in $f(\rm FR)$, is that spirals transform into fast rotators in the relatively low density environments explored by our survey. This is consistent with the close morphological similarity between the two classes of objects indicated in \refsec{sec:morph}. Assuming the current local density is related to the one in which the spiral evolved into a fast-rotator, the relation for $f(\rm Sp)$ down to the most isolated environments, excludes mechanisms related to the cluster environment, like e.g.\ the ram-pressure stripping of the gas due to the inter-galactic medium (IGM) \citep{Spitzer1951,Gunn1972,Abadi1999}, as the dominant way to transform spirals into fast rotators in the low-density regime $\Sigma_{3}\la50$  Mpc$^{-2}$.
Moreover, the fact that galaxy morphology is more closely related to $\Sigma_{3}$ than $\Sigma_{10}$, confirms that, the process which produces the morphological segregation acts on small groups rather than cluster scales.
Given that galaxies are thought to fall into clusters in small groups, the morphological segregation may have happened already before the galaxies enter the cluster and this may explains the existence of a smooth $T-\Sigma$ relation with $\Sigma_{3}$ even inside the Virgo cluster.

Possible processes \citep[see][for a review]{Boselli2006} that can transform spiral galaxies into fast-rotators in the field, or reduce spiral formation via cold accretion \citep{Dekel2009} in the first place, include quenching produced by galaxy harassment \citep{Moore1996,Moore1999}, or gravitational heating by minor merging, which may suppress star formation \citep{Khochfar2008,Johansson2009}. Additional reduction of star formation will result from morphological quenching \citep{Martig2009} caused by the thickening of the disks and growth of the bulges during close encounters, minor mergers, or secular evolution \citep{Kormendy2004,Debattista2006}. The negative trend between the amount of gas and the bulge fraction predicted by all these process is in general agreement with the observations (fig.~4 of Paper~I; \citealt{Dressler1980}), as is the presence of thick disks in S0s \citep{Burstein1979}, which are expected to form via dynamical disturbances \citep{Read2008}.

The change in slope and the sharp decrease in the spiral fraction above $\Sigma_{3}\ga50$  Mpc$^{-2}$ indicates that a different process starts playing a role at the highest densities which are reached for our sample only towards the Virgo core ($R\la0.5$ Mpc).
In that density regime the warm IGM of the Virgo cluster as traced by the X-ray observations \citep{Bohringer1994} starts ram-pressure stripping spiral galaxies of their gas \citep{Giovanelli1985,diSeregoAlighieri2007,Morganti2006} and prevents further cold accretion \citep{Oosterloo2010}, thus rapidly transforming the spiral galaxies into red and passive fast rotators. This is qualitatively in agreement with numerical simulations \citep{Tonnesen2007} and with detailed predictions computed for a sample of eight clusters, as part of the LoCuSS survey, indicating that ram-stripping becomes important inside $r\la r_{200}/3$  (\citealt{Smith2010}, see also \citealt{Moran2007}), where for Virgo the virial radius is $r_{200}\approx1.6$ Mpc \citep{Hoffman1980,McLaughlin1999,Cote2001}. Ram stripping would explain the existence of very thin and dynamically cold fast rotators which are unlikely to have suffered from significant gravitational disturbances (\reffig{fig:morphology_scheme}; see also \citealt{vanderWel2010}). The importance of the IGM near the cluster core ($R<0.5$ Mpc) is also evident in the studies of \citet{Chung2007,Chung2009}, who find only truncated and displaced \hi\ disks in spirals in that region, while more extended and regular disks reside further out. The steepening of the $T-\Sigma$ relation then indicates the addition of the IGM effect on top of the pre-processing of the galaxies morphology that likely happened in groups before the galaxies entered the cluster. Our conclusions are entirely consistent with the ones by \citet{Moran2007}, based on the analysis of the stellar population of two clusters at intermediate redshift. The role of the environment and the IGM on the gaseous components of galaxies is discussed in detail in other papers of this series (Paper~IV; Serra et al. in preparation).

The sharp increase in the fraction of slow rotators inside the Virgo core, as opposed to its outskirts, may be related to the large number of frequent close encounter and minor mergers among the numerous gas-poor galaxies, combined with the lack of cold accretion, and lack of recycling of gas lost during stellar evolution \citep{Leitner2010}, due to the hot IGM. Gas poor mergers in fact generally produce slow rotators \citep[e.g.][]{Jesseit2009,Bois2010,Hoffman2010} and large numbers of minor dry mergers \citep{Bournaud2007} could explain the observed nearly round shape of slow rotators, both inside and outside Virgo. The dominance of massive slow rotators in the Virgo core may also be due to mass segregation that brought them there from larger radii via dynamical friction, towards the bottom of the cluster potential well.

Paper~VIII presents a semi-analytic model for galaxy formation within the hierarchical $\Lambda$CDM scenario. It demonstrates that making the assumption, empirically discussed in \refsec{sec:morph}, that all fast rotators are characterized by the presence of stellar disks with a range of mass fractions (\reffig{fig:morphology_scheme_1}), and by quantitatively treating many of the effects discussed in this section, one can reproduce in detail the observed fractions of fast and slow rotators as a function of luminosity (Paper~III). A key additional test to future models is provided by the observed environmental dependencies presented in this paper.

\section{Summary}

In Paper~I we introduced the volume-limited parent sample of 871 galaxies from which we extracted the \atl\ sample of 260 ETGs. In Paper~II and III we illustrated the kinematic classification of the ETGs into fast and slow rotators, according to their stellar angular momentum parameter $\lambda_R$. In this paper we looked at the morphology of the galaxies. We gave an overview of the limitations of the classic \citet{Hubble1936} tuning-fork diagram, and the usefulness of a scheme similar to the one proposed by \citet{vandenBergh1976}, to properly understand the morphological content of the ETGs of our sample. These are composed of two classes of objects: (i) the slow-rotators which are consistent with being genuinely elliptical-like objects, with intrinsic ellipticity $\varepsilon\la0.4$ and (ii) the fast-rotators which are generally flatter than $\varepsilon\ga0.4$ and are morphologically similar to spiral galaxies with the dust removed, or in some cases to flat ellipticals with disky isophotes, and span the same full range of bulge sizes of spirals. We presented a revised scheme to illustrate the morphology of nearby galaxies, which overcomes the limitations of the tuning-fork diagram.

Only one third (34\%) of the morphologically classified ellipticals are genuine spheroidal slow-rotators (even less considering counter-rotating disks like systems like NGC~4550), while the rest are misclassified lenticular-like systems. Motivated by these findings, we study for the first time the $T-\Sigma$ relation \citep{Dressler1980} using a robust kinematic classification for ETGs, as refined and measured for our \atl\ sample. This method separates ETGs into fast and slow rotators, instead of lenticular and elliptical galaxies, in a way that is nearly insensitive to projection effects.
With respect to the numerous previous studies of the $T-\Sigma$ relation based on morphology alone we find the following new results:
\begin{enumerate}
\item The slow-rotator elliptical-like galaxies are nearly absent at the lowest densities ($f(\rm SR)\la2\%$) and generally constitute a small ($f(\rm SR)\approx4\%$) contribution by number to the entire galaxy population in all environments, with the exception of the Virgo core ($f(\rm SR)\approx20\%$). The process that forms slow rotators must be very inefficient in the field and in small groups;
\item There is a decrease of the spirals fraction $f(\rm Sp)$ and a corresponding increase of the fast rotators fraction $f(\rm FR)$, which is well described by a log-linear relation over nearly four orders of magnitude of the surface density down to $\Sigma_3\approx0.01$ Mpc$^2$, with the fractions changing by 10\% per decade in $\Sigma_3$. The fact that the rate of transformation continues unchanged at the lowest densities, excludes processes related to the cluster environment, like e.g.\ ram-pressure stripping as significant contributors of this segregation in low-density environments.
\item When using a less-local density estimator $\Sigma_{10}$ the linear relation becomes shallower and less well-defined. The average morphology of a galaxy depends on the distance to the third nearest galaxies, independently on the environment at larger distances or on whether the galaxy belongs to Virgo or not. The observed $T-\Sigma$ relation is driven by group-scale and not cluster-scale effects;
\item The $T-\Sigma$ relation shows a break and dramatically steepens inside the densest core of the Virgo cluster, and there the fraction of slow-rotators also dramatically increases. A different process must start acting inside the densest environment of the Virgo cluster. There the presence of hot gas and the ram-pressure stripping can be a viable mechanism for the transformation of spirals into anemic spirals and later into fast-rotators by fading, due to the lack of cold accretion. The dry mergers will produce slow rotators and the large number of minor ones could explain their generally round shape.
\end{enumerate}

The nearby Universe does not include very high density environments. For this reason it would be important in the future to extend the study of the kinematic $T-\Sigma$ relation to denser environments than discussed in this paper and to increasingly higher redshifts as currently done for the classic $T-\Sigma$ relation. This would allow one to better understand the effect of extreme environments on the formation of the fast and slow rotator ETGs and their time evolution.

\section*{acknowledgements}

MC acknowledges support from a STFC Advanced Fellowship PP/D005574/1 and a Royal Society University Research Fellowship.
This work was supported by the rolling grants `Astrophysics at Oxford' PP/E001114/1 and ST/H002456/1 and visitors grants PPA/V/S/2002/00553, PP/E001564/1 and ST/H504862/1 from the UK Research Councils. RLD acknowledges travel and computer grants from Christ Church, Oxford and support from the Royal Society in the form of a Wolfson Merit Award 502011.K502/jd. RLD also acknowledges the support of the ESO Visitor Programme which funded a 3 month stay in 2010.
SK acknowledges support from the the Royal Society Joint Projects Grant JP0869822.
RMcD is supported by the Gemini Observatory, which is operated by the Association of Universities for Research in Astronomy, Inc., on behalf of the international Gemini partnership of Argentina, Australia, Brazil, Canada, Chile, the United Kingdom, and the United States of America.
TN and MBois acknowledge support from the DFG Cluster of Excellence `Origin and Structure of the Universe'.
MS acknowledges support from a STFC Advanced Fellowship ST/F009186/1.
NS and TD acknowledge support from an STFC studentship.
The authors acknowledge financial support from ESO.
MC acknowledges useful comments on the classification scheme from Ryan Houghton, Michael Williams and Christi Warner.
The SAURON observations were obtained at the William Herschel Telescope, operated by the Isaac Newton Group in the Spanish Observatorio del Roque de los Muchachos of the Instituto de Astrofisica de Canarias.
This research has made use of the NASA/IPAC Extragalactic Database (NED) which is operated by the Jet Propulsion Laboratory, California Institute of Technology, under contract with the National Aeronautics and Space Administration. We acknowledge the usage of the HyperLeda database (http://leda.univ-lyon1.fr). Funding for the SDSS and SDSS-II was provided by the Alfred P. Sloan Foundation, the Participating Institutions, the National Science Foundation, the U.S. Department of Energy, the National Aeronautics and Space Administration, the Japanese Monbukagakusho, the Max Planck Society, and the Higher Education Funding Council for England. The SDSS was managed by the Astrophysical Research Consortium for the Participating Institutions. This publication makes use of data products from the Two Micron All Sky Survey, which is a joint project of the University of Massachusetts and the Infrared Processing and Analysis Center/California Institute of Technology, funded by the National Aeronautics and Space Administration and the National Science Foundation.

\bibliographystyle{mn2e}


\clearpage
\centering
\begin{deluxetable}{rrrrrrr}
\tablewidth{0pt}
\setlength{\tabcolsep}{13pt}
\tabletypesize{\small}
\tablecaption{The local galaxy density for the \atl\ sample of 260 early-type galaxies\label{tab:atlas3d_sample}}
\tablehead{
 \colhead{Galaxy} &
 \colhead{$\log\rho_{10}$} &
 \colhead{$\log\Sigma_{10}$} &
 \colhead{$\log\Sigma_3$} &
 \colhead{$\log\nu_{10}$} &
 \colhead{$\log I_{10}$} &
 \colhead{$\log I_{3}$}\\
 \colhead{} &
 \colhead{(Mpc$^{-3}$)} &
 \colhead{(Mpc$^{-2}$)} &
 \colhead{(Mpc$^{-2}$)} &
 \colhead{($L_{\sun K}$Mpc$^{-3}$)} &
 \colhead{($L_{\sun K}$Mpc$^{-2}$)} &
 \colhead{($L_{\sun K}$Mpc$^{-2}$)} \\
 \colhead{(1)} &
 \colhead{(2)} &
 \colhead{(3)} &
 \colhead{(4)} &
 \colhead{(5)} &
 \colhead{(6)} &
 \colhead{(7)}
}
\startdata
     IC0560 &  -1.91 &  -0.76 &   0.05 &   8.57 &   9.74 &  10.43 \\
     IC0598 &  -2.31 &  -1.33 &  -1.71 &   8.27 &   9.28 &   8.91 \\
     IC0676 &  -1.41 &  -0.62 &  -0.62 &   9.19 &   9.96 &   9.79 \\
     IC0719 &  -1.91 &  -1.02 &  -1.21 &   8.93 &   9.73 &   9.17 \\
     IC0782 &  -0.65 &   0.28 &   0.96 &  10.11 &  11.00 &  11.67 \\
     IC1024 &  -0.85 &   0.17 &  -0.10 &   9.82 &  10.91 &  10.53 \\
     IC3631 &  -2.05 &  -1.10 &  -1.06 &   8.89 &   9.58 &   9.80 \\
    NGC0448 &  -1.79 &  -1.01 &  -1.08 &   9.09 &   9.60 &   9.54 \\
    NGC0474 &  -1.78 &  -0.55 &  -0.04 &   9.02 &  10.36 &  11.27 \\
    NGC0502 &  -1.46 &  -0.25 &   1.10 &   9.06 &  10.63 &  11.58 \\
    NGC0509 &  -1.28 &  -0.25 &   1.27 &   9.56 &  10.64 &  12.11 \\
    NGC0516 &  -1.42 &   0.10 &   1.71 &   9.41 &  10.74 &  12.56 \\
    NGC0524 &  -2.22 &   0.01 &   1.80 &   8.44 &  10.79 &  12.32 \\
    NGC0525 &  -1.48 &  -0.57 &   1.27 &   9.38 &  10.41 &  12.06 \\
    NGC0661 &  -2.51 &  -0.89 &   0.07 &   8.39 &  10.02 &  11.05 \\
    NGC0680 &  -1.77 &  -1.03 &   1.47 &   9.09 &   9.71 &  12.15 \\
    NGC0770 &  -2.03 &  -1.31 &  -0.71 &   8.89 &   9.62 &  10.49 \\
    NGC0821 &  -2.12 &  -1.32 &  -1.02 &   8.90 &   9.30 &   9.51 \\
    NGC0936 &  -1.59 &  -0.48 &  -0.56 &   9.34 &  10.35 &  10.60 \\
    NGC1023 &  -2.42 &  -1.45 &  -0.16 &   8.09 &   9.24 &  10.30 \\
    NGC1121 &  -1.98 &  -0.95 &  -0.80 &   8.70 &   9.74 &  10.11 \\
    NGC1222 &  -1.90 &  -1.01 &  -0.53 &   8.60 &   9.58 &   9.88 \\
    NGC1248 &  -1.97 &  -1.03 &  -0.25 &   8.57 &   9.52 &  10.05 \\
    NGC1266 &  -2.11 &  -1.15 &  -0.43 &   8.40 &   9.35 &   9.94 \\
    NGC1289 &  -2.12 &  -1.17 &  -1.10 &   8.46 &   9.43 &   9.44 \\
    NGC1665 &  -1.89 &  -0.96 &   0.38 &   8.47 &   9.40 &  10.65 \\
    NGC2481 &  -2.29 &  -1.18 &  -1.08 &   8.20 &   9.40 &   9.40 \\
    NGC2549 &  -1.99 &  -0.71 &  -0.77 &   8.65 &   9.70 &   9.71 \\
    NGC2577 &  -2.31 &  -1.30 &  -1.03 &   8.15 &   9.26 &   9.52 \\
    NGC2592 &  -2.29 &  -1.02 &  -0.79 &   8.29 &   9.46 &   9.73 \\
    NGC2594 &  -2.18 &  -1.44 &  -1.32 &   8.31 &   9.08 &   9.18 \\
    NGC2679 &  -1.99 &  -0.84 &  -0.78 &   8.49 &   9.59 &   9.56 \\
    NGC2685 &  -1.82 &  -1.12 &  -0.90 &   8.98 &   9.45 &   9.90 \\
    NGC2695 &  -2.23 &  -1.08 &   1.42 &   8.43 &   9.68 &  11.88 \\
    NGC2698 &  -1.94 &  -0.96 &   1.84 &   8.69 &   9.82 &  12.58 \\
    NGC2699 &  -1.91 &  -0.93 &   1.87 &   8.74 &   9.85 &  12.66 \\
    NGC2764 &  -2.01 &  -1.52 &  -0.91 &   8.47 &   8.86 &   9.50 \\
    NGC2768 &  -1.81 &  -0.69 &  -0.07 &   8.85 &   9.95 &  10.43 \\
    NGC2778 &  -1.98 &  -0.92 &  -0.05 &   8.56 &   9.48 &  10.26 \\
    NGC2824 &  -2.08 &  -1.67 &  -0.79 &   8.41 &   8.85 &   9.61 \\
    NGC2852 &  -1.88 &  -0.82 &  -0.91 &   8.67 &   9.74 &   9.46 \\
    NGC2859 &  -1.69 &  -0.61 &  -0.72 &   8.70 &   9.83 &   9.78 \\
    NGC2880 &  -1.73 &  -0.64 &  -0.19 &   9.06 &  10.05 &  10.61 \\
    NGC2950 &  -1.77 &  -0.13 &  -0.05 &   8.91 &  10.52 &  10.77 \\
    NGC2962 &  -2.15 &  -0.47 &  -0.71 &   8.27 &   9.96 &   9.62 \\
    NGC2974 &  -2.05 &  -0.52 &  -0.49 &   8.69 &   9.91 &   9.89 \\
    NGC3032 &  -1.52 &  -0.46 &  -0.23 &   8.99 &  10.16 &  10.43 \\
    NGC3073 &  -2.06 &  -1.06 &  -0.75 &   8.36 &   9.60 &   9.94 \\
    NGC3098 &  -1.18 &  -0.38 &  -0.56 &   9.41 &  10.37 &  10.18 \\
    NGC3156 &  -1.87 &  -1.05 &  -0.58 &   8.79 &   9.55 &  10.31 \\
    NGC3182 &  -1.92 &  -1.06 &  -0.96 &   8.51 &   9.47 &   9.41 \\
    NGC3193 &  -2.43 &  -0.29 &   0.56 &   8.23 &  10.32 &  11.32 \\
    NGC3226 &  -1.11 &   0.16 &   0.16 &   9.49 &  10.86 &  10.77 \\
    NGC3230 &  -1.78 &  -0.75 &  -0.76 &   8.68 &   9.83 &   9.69 \\
    NGC3245 &  -1.36 &  -0.27 &  -0.44 &   8.93 &  10.25 &  10.22 \\
    NGC3248 &  -1.12 &  -0.28 &  -0.03 &   9.57 &  10.47 &  10.80 \\
    NGC3301 &  -1.03 &  -0.06 &  -0.16 &   9.57 &  10.68 &   9.97 \\
    NGC3377 &  -1.25 &  -0.14 &   0.64 &   9.41 &  10.44 &  11.23 \\
    NGC3379 &  -1.13 &  -0.08 &   1.25 &   9.51 &  10.60 &  11.88 \\
    NGC3384 &  -1.27 &   0.03 &   1.12 &   9.37 &  10.64 &  11.69 \\
    NGC3400 &  -1.11 &  -0.08 &   0.38 &   9.41 &  10.45 &  10.93 \\
    NGC3412 &  -1.27 &   0.18 &   1.32 &   9.41 &  10.80 &  12.01 \\
    NGC3414 &  -1.18 &  -0.16 &   0.71 &   9.23 &  10.26 &  10.80 \\
    NGC3457 &  -1.01 &  -0.17 &  -0.26 &   9.61 &  10.34 &   9.97 \\
    NGC3458 &  -1.50 &  -0.22 &  -0.48 &   9.09 &  10.35 &  10.16 \\
    NGC3489 &  -1.38 &  -0.26 &   0.35 &   9.28 &  10.31 &  10.88 \\
    NGC3499 &  -1.53 &  -0.35 &  -0.38 &   9.06 &  10.18 &  10.28 \\
    NGC3522 &  -1.33 &  -0.30 &  -0.43 &   9.32 &  10.19 &  10.16 \\
    NGC3530 &  -1.56 &  -0.27 &   0.04 &   9.07 &  10.37 &  10.74 \\
    NGC3595 &  -1.98 &  -1.05 &  -1.15 &   8.88 &   9.71 &   9.57 \\
    NGC3599 &  -1.19 &  -0.45 &  -0.43 &   9.42 &  10.31 &  10.44 \\
    NGC3605 &  -1.18 &  -0.72 &  -0.50 &   9.44 &   9.86 &  10.05 \\
    NGC3607 &  -0.92 &  -0.95 &  -0.88 &   9.55 &   9.59 &   9.30 \\
    NGC3608 &  -0.96 &  -0.12 &   0.15 &   9.63 &  10.28 &  10.62 \\
    NGC3610 &  -1.04 &  -0.20 &   0.82 &   9.63 &  10.39 &  11.36 \\
    NGC3613 &  -1.58 &  -0.76 &   0.29 &   8.77 &   9.63 &  10.60 \\
    NGC3619 &  -1.38 &  -0.21 &   0.34 &   9.27 &  10.30 &  11.12 \\
    NGC3626 &  -1.36 &  -0.37 &   0.36 &   9.22 &  10.07 &  10.90 \\
    NGC3630 &  -1.96 &  -0.99 &   0.22 &   8.62 &   9.83 &  11.06 \\
    NGC3640 &  -2.02 &  -1.08 &  -0.93 &   8.42 &   9.60 &   9.62 \\
    NGC3641 &  -1.99 &  -1.04 &  -0.92 &   8.62 &   9.75 &   9.63 \\
    NGC3648 &  -1.96 &  -1.31 &  -0.16 &   8.78 &   9.40 &  10.77 \\
    NGC3658 &  -1.96 &  -1.22 &   0.21 &   8.71 &   9.51 &  11.12 \\
    NGC3665 &  -1.94 &  -1.24 &   0.30 &   8.74 &   9.35 &  10.82 \\
    NGC3674 &  -1.75 &  -0.94 &   0.15 &   8.88 &   9.64 &  10.77 \\
    NGC3694 &  -1.97 &  -1.25 &  -0.45 &   8.77 &   9.40 &   9.64 \\
    NGC3757 &  -0.88 &  -0.06 &   0.21 &   9.75 &  10.65 &  10.37 \\
    NGC3796 &  -0.85 &  -0.15 &   0.21 &   9.99 &  10.54 &  10.87 \\
    NGC3838 &  -0.84 &   0.12 &   0.16 &   9.88 &  10.76 &  10.73 \\
    NGC3941 &  -1.20 &  -0.04 &  -0.26 &   9.20 &  10.65 &  10.42 \\
    NGC3945 &  -0.79 &  -0.04 &   0.25 &   9.94 &  10.53 &  11.02 \\
    NGC3998 &  -1.01 &   0.52 &   0.94 &   9.49 &  11.25 &  11.57 \\
    NGC4026 &  -0.74 &   0.55 &   0.77 &   9.83 &  11.32 &  11.51 \\
    NGC4036 &  -0.69 &  -0.09 &  -0.06 &   9.97 &  10.56 &  10.68 \\
    NGC4078 &  -1.32 &  -0.50 &  -0.73 &   9.72 &  10.22 &   9.82 \\
    NGC4111 &  -0.61 &   0.28 &   0.79 &   9.91 &  10.89 &  11.65 \\
    NGC4119 &  -0.05 &   0.42 &   0.06 &  10.49 &  10.88 &  10.03 \\
    NGC4143 &  -0.66 &   0.22 &   0.33 &   9.81 &  10.78 &  10.87 \\
    NGC4150 &  -1.18 &  -2.03 &  -1.24 &   9.49 &   8.74 &   9.39 \\
    NGC4168 &  -1.93 &  -1.05 &  -1.12 &   9.07 &  10.07 &  10.21 \\
    NGC4179 &  -0.34 &   0.07 &  -0.12 &  10.14 &  10.54 &  10.51 \\
    NGC4191 &  -1.03 &   0.06 &  -0.12 &   9.67 &  10.75 &  10.80 \\
    NGC4203 &  -1.02 &  -0.47 &  -0.25 &   9.54 &  10.11 &  10.20 \\
    NGC4215 &  -1.59 &  -0.76 &   0.16 &   9.24 &   9.83 &  10.91 \\
    NGC4233 &  -1.26 &   0.19 &   0.69 &   9.54 &  10.86 &  11.61 \\
    NGC4249 &  -0.81 &   0.28 &   0.98 &   9.90 &  11.01 &  11.32 \\
    NGC4251 &  -1.37 &  -0.81 &  -0.47 &   9.59 &  10.03 &   9.90 \\
    NGC4255 &  -1.45 &  -0.81 &   0.59 &   9.48 &   9.85 &  11.24 \\
    NGC4259 &  -0.57 &   0.32 &   1.25 &  10.25 &  11.05 &  11.99 \\
    NGC4261 &  -1.55 &  -0.90 &  -0.46 &   9.13 &  10.15 &  10.65 \\
    NGC4262 &   0.46 &   1.08 &   1.14 &  11.19 &  11.69 &  11.45 \\
    NGC4264 &  -0.61 &   0.37 &   1.00 &  10.16 &  11.14 &  11.49 \\
    NGC4267 &   0.57 &   1.17 &   0.92 &  11.10 &  12.05 &  11.68 \\
    NGC4268 &  -1.53 &  -0.74 &   1.03 &   9.27 &   9.92 &  11.68 \\
    NGC4270 &  -0.85 &   0.43 &   1.23 &   9.83 &  11.01 &  11.67 \\
    NGC4278 &  -1.13 &  -0.85 &   0.36 &   9.72 &  10.01 &  11.55 \\
    NGC4281 &  -1.76 &   0.52 &   1.34 &   9.24 &  11.21 &  11.53 \\
    NGC4283 &  -1.07 &  -0.56 &   0.15 &   9.83 &  10.07 &  10.77 \\
    NGC4324 &   0.43 &   0.88 &   1.21 &  10.82 &  11.28 &  11.86 \\
    NGC4339 &   0.46 &   0.83 &   1.11 &  10.85 &  11.63 &  11.35 \\
    NGC4340 &   0.11 &   1.09 &   1.15 &  10.91 &  11.87 &  11.42 \\
    NGC4342 &   0.52 &   0.94 &   1.09 &  11.06 &  11.69 &  11.27 \\
    NGC4346 &  -0.69 &   0.35 &   0.85 &   9.90 &  10.88 &  11.41 \\
    NGC4350 &   0.31 &   1.24 &   1.40 &  11.02 &  12.02 &  11.64 \\
    NGC4365 &  -1.74 &  -0.47 &   0.18 &   9.21 &  10.59 &  11.28 \\
    NGC4371 &   1.20 &   1.34 &   1.53 &  12.14 &  12.16 &  11.66 \\
    NGC4374 &   0.10 &   1.34 &   1.89 &  10.50 &  12.06 &  12.82 \\
    NGC4377 &   0.54 &   1.24 &   1.14 &  10.95 &  11.80 &  11.56 \\
    NGC4379 &   0.69 &   1.47 &   1.52 &  11.44 &  11.98 &  12.22 \\
    NGC4382 &  -0.12 &   0.74 &   0.75 &  10.34 &  11.38 &  11.59 \\
    NGC4387 &   0.45 &   1.41 &   1.98 &  11.25 &  12.28 &  13.17 \\
    NGC4406 &   1.18 &   1.56 &   2.14 &  11.96 &  12.31 &  13.10 \\
    NGC4417 &   0.75 &   1.13 &   1.58 &  11.12 &  11.99 &  12.10 \\
    NGC4425 &   1.25 &   1.80 &   1.95 &  12.21 &  12.65 &  13.02 \\
    NGC4429 &   1.08 &   1.30 &   1.19 &  11.82 &  11.70 &  11.56 \\
    NGC4434 &  -1.54 &  -0.37 &  -0.19 &   9.47 &  10.79 &  11.12 \\
    NGC4435 &   1.26 &   1.78 &   1.95 &  11.98 &  12.62 &  12.54 \\
    NGC4442 &   0.56 &   1.19 &   1.44 &  10.87 &  11.68 &  11.80 \\
    NGC4452 &   0.85 &   1.43 &   1.43 &  11.43 &  12.19 &  11.68 \\
    NGC4458 &   1.03 &   1.72 &   1.97 &  11.83 &  12.54 &  12.73 \\
    NGC4459 &   0.79 &   1.51 &   1.55 &  11.45 &  12.23 &  12.25 \\
    NGC4461 &   1.04 &   1.71 &   1.97 &  11.95 &  12.50 &  12.66 \\
    NGC4472 &   0.72 &   1.07 &   1.27 &  11.21 &  11.59 &  11.56 \\
    NGC4473 &   0.65 &   1.65 &   2.15 &  11.40 &  12.42 &  12.72 \\
    NGC4474 &   0.79 &   1.47 &   1.53 &  11.65 &  12.25 &  12.52 \\
    NGC4476 &   0.65 &   1.59 &   2.20 &  11.40 &  12.34 &  13.25 \\
    NGC4477 &   0.88 &   1.48 &   1.79 &  11.70 &  12.21 &  12.51 \\
    NGC4478 &   1.41 &   1.57 &   2.55 &  12.33 &  12.31 &  13.58 \\
    NGC4483 &   0.81 &   1.19 &   1.26 &  11.74 &  12.09 &  11.50 \\
    NGC4486 &   1.13 &   1.52 &   2.38 &  11.86 &  11.96 &  12.60 \\
   NGC4486A &   0.17 &   1.40 &   2.17 &  11.04 &  12.11 &  13.21 \\
    NGC4489 &   0.34 &   1.12 &   0.87 &  11.08 &  11.56 &  11.54 \\
    NGC4494 &  -0.93 &  -0.91 &  -0.26 &  10.03 &   9.82 &  10.82 \\
    NGC4503 &   1.04 &   1.37 &   0.98 &  11.92 &  12.14 &  11.23 \\
    NGC4521 &  -1.29 &   0.01 &   0.98 &   9.33 &  10.64 &  11.41 \\
    NGC4526 &   0.83 &   1.03 &   1.21 &  11.21 &  11.84 &  11.82 \\
    NGC4528 &   0.74 &   1.37 &   1.54 &  11.26 &  12.24 &  12.07 \\
    NGC4546 &  -1.14 &  -0.31 &  -0.43 &   9.56 &  10.66 &  10.40 \\
    NGC4550 &   0.81 &   1.55 &   1.46 &  11.62 &  12.29 &  12.41 \\
    NGC4551 &   0.99 &   1.49 &   1.42 &  11.71 &  12.24 &  12.37 \\
    NGC4552 &   0.91 &   1.38 &   1.51 &  11.54 &  12.23 &  11.70 \\
    NGC4564 &   0.75 &   1.43 &   1.71 &  11.33 &  12.07 &  12.60 \\
    NGC4570 &   0.52 &   0.99 &   0.92 &  11.35 &  11.85 &  11.77 \\
    NGC4578 &   0.85 &   0.98 &   0.82 &  11.50 &  11.47 &  11.33 \\
    NGC4596 &   0.85 &   1.14 &   0.80 &  11.42 &  11.97 &  11.37 \\
    NGC4608 &   0.67 &   1.11 &   0.82 &  11.49 &  11.94 &  11.38 \\
    NGC4612 &   0.51 &   0.71 &   0.50 &  11.06 &  11.30 &  11.34 \\
    NGC4621 &   0.30 &   1.38 &   2.00 &  11.06 &  12.22 &  12.41 \\
    NGC4623 &   0.22 &   0.73 &   0.54 &  11.11 &  11.35 &  11.16 \\
    NGC4624 &  -0.18 &   0.24 &   0.17 &  10.34 &  10.72 &  10.81 \\
    NGC4636 &  -0.70 &   0.05 &   0.08 &   9.95 &  11.08 &  10.81 \\
    NGC4638 &   0.41 &   1.21 &   2.13 &  11.38 &  12.08 &  13.30 \\
    NGC4643 &  -0.29 &   0.07 &   0.22 &  10.23 &  10.54 &  10.86 \\
    NGC4649 &   0.57 &   1.15 &   1.80 &  11.10 &  11.79 &  12.52 \\
    NGC4660 &   0.18 &   1.24 &   1.75 &  10.92 &  12.03 &  12.86 \\
    NGC4684 &  -1.24 &   0.16 &   0.34 &   9.44 &  11.00 &  11.27 \\
    NGC4690 &  -1.63 &  -0.66 &  -0.44 &   8.99 &  10.22 &   9.97 \\
    NGC4694 &   0.62 &   1.00 &   1.02 &  11.50 &  11.80 &  11.52 \\
    NGC4697 &  -1.63 &   0.56 &   0.53 &   9.13 &  11.35 &  10.75 \\
    NGC4710 &   0.09 &   0.45 &   0.42 &  10.77 &  11.08 &  10.72 \\
    NGC4733 &  -0.32 &   0.88 &   1.02 &  10.50 &  11.71 &  11.47 \\
    NGC4753 &  -1.23 &  -0.25 &  -0.01 &   9.58 &  10.48 &  10.23 \\
    NGC4754 &   0.25 &   0.74 &   1.04 &  10.75 &  11.51 &  11.11 \\
    NGC4762 &  -1.61 &  -0.68 &  -0.71 &   9.34 &  10.43 &   9.87 \\
    NGC4803 &  -1.90 &  -0.80 &  -0.28 &   8.59 &   9.52 &  10.10 \\
    NGC5103 &  -2.01 &  -1.03 &  -0.73 &   8.47 &   9.54 &   9.99 \\
    NGC5173 &  -1.72 &  -0.78 &   0.63 &   9.00 &   9.89 &  11.23 \\
    NGC5198 &  -1.63 &  -0.73 &   0.78 &   8.98 &   9.84 &  11.07 \\
    NGC5273 &  -1.74 &  -0.76 &  -0.62 &   9.14 &   9.94 &  10.01 \\
    NGC5308 &  -1.34 &  -0.46 &   0.32 &   9.44 &  10.34 &  11.31 \\
    NGC5322 &  -1.22 &  -0.38 &   0.56 &   9.43 &  10.29 &  11.33 \\
    NGC5342 &  -1.76 &  -0.59 &   0.07 &   9.05 &  10.20 &  11.00 \\
    NGC5353 &  -1.35 &  -0.54 &   2.34 &   9.26 &  10.10 &  12.99 \\
    NGC5355 &  -0.81 &   0.93 &   2.53 &   9.98 &  11.84 &  13.59 \\
    NGC5358 &  -0.63 &   0.92 &   1.98 &  10.08 &  11.84 &  13.05 \\
    NGC5379 &  -1.17 &  -0.22 &   0.63 &   9.67 &  10.63 &  11.83 \\
    NGC5422 &  -0.98 &  -0.27 &   0.65 &   9.85 &  10.45 &  11.43 \\
    NGC5473 &  -1.37 &  -0.27 &   0.85 &   9.19 &  10.26 &  11.49 \\
    NGC5475 &  -1.21 &  -0.29 &   0.71 &   9.65 &  10.56 &  11.46 \\
    NGC5481 &  -1.78 &  -0.07 &   0.23 &   8.74 &  10.59 &  10.87 \\
    NGC5485 &  -1.80 &  -0.05 &   0.74 &   8.76 &  10.57 &  11.56 \\
    NGC5493 &  -1.96 &  -0.59 &   0.06 &   8.64 &  10.01 &  10.48 \\
    NGC5500 &  -1.40 &  -0.22 &  -0.03 &   9.37 &  10.53 &  10.52 \\
    NGC5507 &  -1.46 &  -0.62 &  -0.14 &   9.18 &  10.08 &  10.43 \\
    NGC5557 &  -1.60 &  -0.38 &   0.30 &   8.91 &  10.41 &  10.73 \\
    NGC5574 &  -0.98 &  -0.11 &   1.06 &   9.68 &  10.54 &  11.99 \\
    NGC5576 &  -0.86 &  -0.16 &   1.03 &   9.71 &  10.46 &  11.80 \\
    NGC5582 &  -2.01 &  -1.44 &  -1.12 &   8.59 &   8.97 &   8.91 \\
    NGC5611 &  -2.37 &  -1.13 &  -1.29 &   8.18 &   9.58 &   9.73 \\
    NGC5631 &  -1.58 &  -0.06 &  -0.13 &   9.23 &  10.64 &  10.56 \\
    NGC5638 &  -0.76 &   0.00 &   0.02 &  10.09 &  10.70 &  10.63 \\
    NGC5687 &  -1.66 &  -0.23 &  -0.56 &   8.97 &  10.49 &  10.18 \\
    NGC5770 &  -1.96 &   0.32 &   1.32 &   8.86 &  11.17 &  11.99 \\
    NGC5813 &  -1.64 &  -0.22 &  -0.04 &   9.23 &  10.79 &  10.98 \\
    NGC5831 &  -0.97 &   0.10 &   0.77 &   9.97 &  11.09 &  11.96 \\
    NGC5838 &  -1.45 &  -0.39 &   0.87 &   9.27 &  10.20 &  11.45 \\
    NGC5839 &  -1.44 &  -1.19 &   0.53 &   9.33 &   9.49 &  11.27 \\
    NGC5845 &  -1.08 &   0.02 &   1.32 &   9.88 &  10.82 &  12.35 \\
    NGC5846 &  -1.00 &   0.10 &   0.71 &   9.63 &  10.96 &  11.37 \\
    NGC5854 &  -0.97 &   0.03 &   0.63 &   9.97 &  10.92 &  11.62 \\
    NGC5864 &  -1.33 &  -0.30 &   0.11 &   9.67 &  10.64 &  11.05 \\
    NGC5866 &  -2.12 &  -1.17 &  -0.73 &   8.58 &   9.43 &   9.88 \\
    NGC5869 &  -1.09 &  -0.58 &  -0.12 &   9.71 &  10.22 &  10.87 \\
    NGC6010 &  -1.99 &  -1.07 &  -0.66 &   8.48 &   9.34 &   9.74 \\
    NGC6014 &  -2.27 &  -1.73 &  -1.84 &   8.15 &   8.92 &   9.00 \\
    NGC6017 &  -1.81 &  -0.78 &  -1.01 &   8.68 &   9.71 &   9.59 \\
    NGC6149 &  -2.55 &  -1.86 &  -1.36 &   8.02 &   8.70 &   9.25 \\
    NGC6278 &  -2.52 &  -1.47 &  -1.36 &   8.01 &   9.18 &   9.23 \\
    NGC6547 &  -2.20 &  -1.46 &  -1.44 &   8.40 &   9.10 &   9.17 \\
    NGC6548 &  -2.62 &  -1.42 &  -0.38 &   8.02 &   9.21 &  10.33 \\
    NGC6703 &  -2.92 &  -1.57 &  -1.33 &   7.73 &   9.05 &   9.39 \\
    NGC6798 &  -2.68 &  -1.63 &  -1.40 &   7.99 &   9.07 &   9.25 \\
    NGC7280 &  -2.13 &  -1.58 &  -1.12 &   8.43 &   8.89 &   9.25 \\
    NGC7332 &  -2.34 &  -1.91 &  -1.20 &   8.29 &   8.68 &   9.53 \\
    NGC7454 &  -1.99 &  -1.34 &  -0.35 &   8.61 &   9.14 &  10.11 \\
    NGC7457 &  -2.44 &  -1.37 &  -1.23 &   8.21 &   9.30 &   9.70 \\
    NGC7465 &  -1.97 &  -1.53 &  -0.33 &   8.54 &   8.89 &  10.24 \\
    NGC7693 &  -2.09 &  -1.45 &  -0.11 &   8.58 &   9.31 &  10.29 \\
    NGC7710 &  -2.13 &  -1.55 &  -0.03 &   8.56 &   9.14 &  10.09 \\
  PGC016060 &  -1.97 &  -0.95 &  -0.05 &   8.42 &   9.43 &  10.38 \\
  PGC028887 &  -2.03 &  -0.87 &  -0.44 &   8.46 &   9.60 &   9.75 \\
  PGC029321 &  -2.00 &  -0.96 &  -0.20 &   8.48 &   9.57 &  10.13 \\
  PGC035754 &  -2.20 &  -1.23 &  -0.14 &   8.55 &   9.30 &  10.55 \\
  PGC042549 &  -1.36 &  -0.51 &   0.84 &   9.17 &  10.32 &  11.99 \\
  PGC044433 &  -2.14 &  -1.16 &  -1.05 &   8.40 &   9.75 &   9.28 \\
  PGC050395 &  -1.70 &  -0.70 &  -0.84 &   8.92 &  10.06 &   9.88 \\
  PGC051753 &  -1.58 &  -0.69 &  -0.55 &   9.04 &  10.06 &   9.89 \\
  PGC054452 &  -1.41 &  -0.57 &  -0.16 &   9.46 &  10.41 &  10.52 \\
  PGC056772 &  -2.60 &  -1.83 &  -1.81 &   7.87 &   8.94 &   8.59 \\
  PGC058114 &  -2.31 &  -1.34 &  -1.04 &   8.22 &   9.45 &   9.38 \\
  PGC061468 &  -2.35 &  -1.78 &  -0.86 &   8.30 &   8.93 &   9.92 \\
  PGC071531 &  -2.06 &  -1.45 &  -1.21 &   8.63 &   9.04 &   9.22 \\
  PGC170172 &  -2.20 &  -1.36 &  -1.37 &   8.65 &   9.38 &   8.67 \\
   UGC03960 &  -2.36 &  -1.10 &  -0.98 &   8.23 &   9.51 &   9.77 \\
   UGC04551 &  -2.24 &  -1.27 &  -1.45 &   8.15 &   9.26 &   9.06 \\
   UGC05408 &  -2.09 &  -1.14 &   0.51 &   8.26 &   9.18 &  10.93 \\
   UGC06062 &  -2.18 &  -1.38 &  -1.29 &   8.50 &   9.11 &   9.22 \\
   UGC06176 &  -2.41 &  -1.35 &  -1.58 &   8.32 &   9.30 &   9.06 \\
   UGC08876 &  -1.43 &  -0.45 &  -0.48 &   9.45 &  10.16 &  10.10 \\
   UGC09519 &  -2.45 &  -1.48 &  -1.58 &   8.19 &   8.96 &   8.75 \\
\enddata
\tablecomments{
Column (1): The Name is the principal designation from LEDA, which is used as standard designation for the \atl\ survey.
Column (2): Mean density of galaxies inside a sphere centered on the galaxy and containing the 10 nearest neighbors. All density estimators only include galaxies brighter than $M_K<-21.5$ mag.
Column (3): Mean surface density of galaxies inside a cylinder of height $h=600$ \kms\ (i.e.\ $\Delta V_{\rm hel}<300$ \kms) centered on the galaxy which contains the 10 nearest neighbors. For galaxies in Virgo the cylinder includes all the cluster galaxies along the line-of-sight.
Column (4): Same as in column 3, but using the three nearest neighbors.
Column (5): Mean $K$-band luminosity density of galaxies inside the sphere defined in column 2. The galaxy luminosity is defined as $L_K \equiv 10^{-0.4 (M_K - M_{\sun K})}$, where the solar absolute luminosity $M_{\sun K}=3.29$ mag \citep{Blanton2007}.
Column (6): Mean $K$-band luminosity surface density of galaxies inside the cylinder defined in column 3.
Column (7): same as in column 6, but using the three nearest neighbors.
}
\end{deluxetable}

\clearpage
\begin{deluxetable}{rrrrrrr}
\tablewidth{0pt}
\setlength{\tabcolsep}{13pt}
\tabletypesize{\small}
\tablecaption{The local galaxy density for the 611 spiral galaxies in the \atl\ parent sample.\label{tab:atlas3d_spirals}}
\tablehead{
 \colhead{Galaxy} &
 \colhead{$\log\rho_{10}$} &
 \colhead{$\log\Sigma_{10}$} &
 \colhead{$\log\Sigma_3$} &
 \colhead{$\log\nu_{10}$} &
 \colhead{$\log I_{10}$} &
 \colhead{$\log I_{3}$}\\
 \colhead{} &
 \colhead{(Mpc$^{-3}$)} &
 \colhead{(Mpc$^{-2}$)} &
 \colhead{(Mpc$^{-2}$)} &
 \colhead{($L_{\sun K}$Mpc$^{-3}$)} &
 \colhead{($L_{\sun K}$Mpc$^{-2}$)} &
 \colhead{($L_{\sun K}$Mpc$^{-2}$)} \\
 \colhead{(1)} &
 \colhead{(2)} &
 \colhead{(3)} &
 \colhead{(4)} &
 \colhead{(5)} &
 \colhead{(6)} &
 \colhead{(7)}
}
\startdata
     IC0065 &  -3.12 &  -1.86 &  -1.54 &   7.59 &   8.66 &   8.72 \\
     IC0163 &  -1.68 &  -0.78 &   0.08 &   9.21 &  10.11 &  11.09 \\
     IC0239 &  -2.48 &  -1.56 &  -1.53 &   8.15 &   9.45 &   9.07 \\
     IC0540 &  -1.72 &  -0.82 &   0.65 &   8.82 &   9.72 &  11.24 \\
     IC0591 &  -1.98 &  -0.93 &  -0.11 &   8.50 &   9.53 &  10.21 \\
     IC0610 &  -1.18 &  -0.08 &   0.15 &   9.30 &  10.66 &  10.83 \\
     IC0750 &  -2.29 &  -0.68 &  -0.14 &   8.42 &   9.93 &  10.45 \\
     IC0777 &  -2.27 &  -1.23 &  -1.00 &   8.29 &   9.32 &   9.29 \\
     IC0800 &  -1.87 &  -0.95 &  -0.96 &   9.09 &  10.00 &  10.39 \\
     IC0851 &  -2.31 &  -1.23 &  -1.36 &   8.24 &   9.31 &   9.36 \\
\enddata
\tablecomments{
The meaning of the columns is the same as in Table~\ref{tab:atlas3d_sample}. Only the first 10 rows are shown while the full 611 will be published electronically. Both Table~\ref{tab:atlas3d_sample} and \ref{tab:atlas3d_spirals} are available from our project website http://purl.org/atlas3d.
}
\end{deluxetable}

\label{lastpage}

\end{document}